\documentclass[%
 reprint,
 groupedaddress,
 superscriptaddress,
 amsmath,amssymb,
 aps,
 prd
]{revtex4-1}

\usepackage{graphicx}
\usepackage{epstopdf,units}
\usepackage{caption}
\usepackage{subcaption}
\usepackage[normalem]{ulem}
\usepackage{dcolumn}
\usepackage{bm}
\usepackage{color}
\usepackage[dvipsnames]{xcolor} 
\graphicspath{{Figures/}} 
\usepackage[%
 colorlinks=true,
 citecolor=blue,
 linkcolor=blue,
 urlcolor=blue
]{hyperref}
\usepackage{enumerate}
\usepackage{float}

\begin{document}

\title{Measuring eccentricity in binary black hole inspirals with gravitational waves}

\author{Marcus E. Lower}
 \email{mlower@swin.edu.au}
 \affiliation{Centre for Astrophysics and Supercomputing, Swinburne University of Technology, Hawthorn VIC 3122, Australia}
 \affiliation{OzGrav: The ARC Centre of Excellence for Gravitational-wave Discovery, Hawthorn VIC 3122, Australia}
 \affiliation{Monash Centre for Astrophysics, School of Physics and Astronomy, Monash University, VIC 3800, Australia}

\author{Eric Thrane}
 \email{eric.thrane@monash.edu}
 \affiliation{Monash Centre for Astrophysics, School of Physics and Astronomy, Monash University, VIC 3800, Australia}
 \affiliation{OzGrav: The ARC Centre of Excellence for Gravitational-wave Discovery, Clayton, VIC 3800, Australia}
 
\author{Paul D. Lasky}
 \email{paul.lasky@monash.edu}
 \affiliation{Monash Centre for Astrophysics, School of Physics and Astronomy, Monash University, VIC 3800, Australia}
 \affiliation{OzGrav: The ARC Centre of Excellence for Gravitational-wave Discovery, Clayton, VIC 3800, Australia}
 
\author{Rory Smith}
 \affiliation{Monash Centre for Astrophysics, School of Physics and Astronomy, Monash University, VIC 3800, Australia}
 \affiliation{OzGrav: The ARC Centre of Excellence for Gravitational-wave Discovery, Clayton, VIC 3800, Australia}

\date{\today}

\begin{abstract}
When binary black holes form in the field, it is expected that their orbits typically circularize before coalescence.  
In galactic nuclei and globular clusters, binary black holes can form dynamically.
Recent results suggest that $\approx5\%$ of mergers in globular clusters result from three-body interactions.
These three-body interactions are expected to induce significant orbital eccentricity $\gtrsim 0.1$ when they enter the Advanced LIGO and Virgo band at a gravitational-wave frequency of $\unit[10]{Hz}$.
Measurements of binary black hole eccentricity therefore provide a means for  determining whether or not dynamic formation is the primary channel for producing binary black hole mergers.
We present a framework for performing Bayesian parameter estimation on gravitational-wave observations of eccentric black hole inspirals.
Using this framework, and employing the nonspinning, inspiral-only {\sc EccentricFD} waveform approximant, we determine the minimum detectable eccentricity for an event with masses and distance similar to GW150914.
At design sensitivity, we find that the current generation of advanced observatories will be sensitive to orbital eccentricities of $\gtrsim0.05$ at a gravitational-wave frequency of $\unit[10]{Hz}$, demonstrating that existing detectors can use eccentricity to distinguish between circular field binaries and globular cluster triples.
We compare this result to eccentricity distributions predicted to result from three black hole binary formation channels, showing that measurements of eccentricity could be used to infer the population properties of binary black holes.

\end{abstract}

\pacs{Valid PACS appear here}
\maketitle

\section{\label{sec:intro}Introduction}

Binary black holes (BBH) are among the most extreme objects in the observable Universe, with their existence having been confirmed through the direct observations of gravitational waves by the Advanced Laser Interferometer Gravitational-wave Observatory (aLIGO) \cite{Aasi2015} and Advanced Virgo (AdV) \cite{Acernese2015}. To date, five confirmed BBH mergers have been observed by Advanced LIGO and Virgo \cite{Abbott150914, Abbott151226, Abbott170104, Abbott170608, Abbott170814}, in addition to one strong candidate \cite{AbbottO1Cat}.  The individual black holes of these systems are believed to have formed through either direct stellar collapse \cite{Belczynski2016}, or from high-mass stars undergoing core-collapse supernovae \cite{Heger2003}. The mechanism by which these black holes came to be in binaries is unknown, although a variety of formation scenarios have been proposed.
Recent work investigates how measurements of black hole mass and spin distributions can elucidate the population properties of binary black holes~\cite{Dominik2015, Stevenson2015, Belczynski2016a, Bartos2017b, Kovetz2017, Talbot2017, Wysocki2017, Miyamoto2017, Mandel2017, Zevin2017, Stevenson2017, Stevenson2017a, Barrett2018, Farr2017, Fishbach2017, OShaughnessy2017, Belczynski2017, Wysocki2018, Talbot2018}.
It has also been suggested that the future Laser Interferometer Space Antenna (LISA) will be able to observe nearby stellar-mass BBH during the early inspiral phase. These observations would allow for long-term tracking of BBH orbital properties which can be used to infer the formation mechanism \cite{Breivik2016}, in addition to precise sky localisation prior to detections made by Advanced LIGO and Virgo \cite{Nishizawa2016, Sesana2016}.
In this work, we focus on the measurement of eccentricity imprinted on the gravitational waveform of binary black holes observed by advanced detectors such as LIGO and Virgo.
We show that measuring the eccentricity of binary black holes using Bayesian parameter estimation can be used to test the dynamical formation hypothesis and other nonstandard channels.
Measurements of eccentricity may also provide information about the globular clusters or galactic nuclei in which black hole binaries might form. 

Binary black holes formed as a result of isolated, massive stellar binary evolution are known as ``field binaries".
The stellar progenitors of these systems are predicted to have undergone either a series of common envelope stages \cite{Belczynski2016} or chemically homogeneous stellar evolution \cite{Mandel2016, demink2016}.
See also~\cite{Tagawa2018}'s proposal for fallback-driven mergers.
Field binaries are expected to circularize by the time they enter the band of Advanced LIGO and Virgo so that the eccentricity is completely undetectable~\cite{Peters1964}.

Dynamic formation of BBH is hypothesized to occur within the dense stellar environments found in globular clusters and galactic nuclei, where the black hole population sinks toward the region of highest stellar density due to dynamic friction, before decoupling from the rest of the stellar environment \cite{Spitzer1969}. This results in a dense subsystem of gravitationally interacting black holes \cite{Freitag2006, Morscher2013}.
Binary black hole formation can then occur through a number of dynamic pathways, including various forms of three-body interactions between stars, other BBH or single black holes and other compact objects such as neutron stars or white dwarfs~\cite{Samsing2014, Rodriguez2016ecc, Park2017}.

A new picture of dynamic mergers is beginning to emerge from recent studies of globular clusters, which include proper treatment of general relativistic effects~\cite{Rodriguez2018a, Samsing2018a, Samsing2018b, Samsing2018c}.
While the overall rate of mergers in globular clusters remains highly uncertain, it is apparent that there are three populations of binary black hole mergers in globular clusters: ejected mergers outside the cluster, two-body mergers inside the cluster, and three-body mergers inside the cluster~\cite{Rodriguez2018a, Samsing2018a, Samsing2018b, Samsing2018c}.
Each population is described by a distinct distribution of gravitational-wave frequencies at formation, with ejected mergers forming at $\sim\unit[10^{-5}-10^{-3}]{Hz}$, two-body mergers inside the cluster forming at $\sim\unit[10^{-4}-10^{-2}]{Hz}$, and three-body mergers forming near the observing band of advanced detectors at $\sim\unit[1-100]{Hz}$~\cite{Rodriguez2018a, Samsing2018a}.

Since they form at low gravitational-wave frequencies, ejected mergers and two-body mergers inside the cluster will circularize by the time they enter the band of advanced detectors.
However, recent work~\cite{Samsing2018a, Rodriguez2018a} suggests that $\approx$5\% of globular cluster mergers are a result three-body interactions, which can enter the LIGO and Virgo bands with significant eccentricities.
The prediction of three subpopulations with $\approx$5\% three-body mergers is only weakly dependent on assumptions about the globular cluster such as the velocity dispersion and black hole density~\cite{Rodriguez2018a, Samsing2018a}.
The robustness of this prediction provides an opportunity to test whether dynamical formation within globular clusters is the primary channel for producing binary black hole mergers, as advanced detectors at design sensitivity will be capable of observing $\gtrsim100$ black hole mergers per year.

Detection of a single eccentric binary could provide evidence that dynamical formation (or other nonstandard evolutionary pathways) is a major source of binary black hole mergers, possibly the dominant one.
On the other hand, if advanced detectors see no evidence of eccentricity after a large number of events, it will be possible to infer that dynamical mergers in globular clusters play a subdominant role in the production of binary black holes.
The BBH mergers observed in the first observing run are consistent with no detectable eccentricity \cite{150914prop, 150914modelsys, AbbottO1Cat}. Studies on the BBH mergers seen in the second observing run with eccentric waveform models are yet to be performed \cite{Abbott170104, Abbott170608, Abbott170814}.

Our work improves upon~\cite{Gondan2018}, which estimates the sensitivity of gravitational-wave detectors to eccentricity using a Fisher matrix approximation.
Fisher matrix calculations such as those in~\cite{Gondan2018} are useful for providing back-of-the-envelope estimates of the sensitivity of gravitational-wave detectors to different parameters.
However, there are well-known limitations on what we learn from them~\cite{Vallisneri2008}.
First, Fisher matrix calculations model the likelihood as a covariant Gaussian when it is actually a more complicated distribution.
As a consequence, the uncertainties quoted from Fisher matrix calculations tend to be overly optimistic.
By carrying out Bayesian parameter estimation, we endeavour to derive results that take into account the full complexity of the likelihood function.

Second, the Fisher matrix calculation does not yield actual posterior distributions for physical quantities; it only provides an estimate for what the posterior width should be if were one to carry out Bayesian inference.
Generating posterior distributions in gravitational-wave astronomy is a computationally expensive task.
Gravitational-wave astronomers typically employ low-cost waveform approximants in order to carry out parameter estimation on reasonable timescales.
As a result, there is sometimes a large gulf between Fisher matrix calculations, which require only a few waveform evaluations, and Bayesian parameter estimation, which requires many thousands.

In this paper, we carry out Bayesian parameter estimation with currently available tools and derive posterior distributions for eccentricity.
This is a first step in what is likely to be a long-term effort to develop increasingly sophisticated Bayesian inference for eccentric binaries.

The remainder of this paper is organized as follows.
In Section~\ref{sec:formalism} we outline the statistical framework and demonstrate the construction of posterior distributions for an eccentric GW150914-like event.
We compare two metrics for distinguishing eccentric inspirals from corresponding quasicircular events: an overlap statistic (Section~\ref{subsec:overlap}) and a Bayes factor (Section~\ref{subsec:lnBF}).
We show that the commonly-used overlap technique can significantly overestimate the sensitivity to eccentricity.
We estimate the minimum distinguishable eccentricity using Bayesian methods, and compare to optimistic estimates found via the waveform overlap for an eccentric GW150914-like event in Section~\ref{sec:BFvsO}.
In Section~\ref{sec:populations} we compare the eccentricity sensitivity of LIGO and Virgo to the distribution of eccentricities predicted for three BBH formation channels, including dynamical mergers in globular clusters.

\section{\label{sec:formalism} Bayesian Framework}

We employ Bayesian inference to determine the parameters describing sources of gravitational waves $\vec{\theta}$ from strain data $h$.
The resulting waveforms from the inspiral of BBH systems on quasielliptical orbits is described by a 17-dimensional parameter space, including the black hole masses $\{m_{1}, m_{2}\}$, spin vectors $\{\vec{S}_{1}, \vec{S}_{2}\}$, orbital eccentricity and argument of periapsis $\{e, \omega\}$, and seven other parameters encoding the orientation and position of the binary relative to the detector.
Parameter estimation of gravitational-wave signals is performed using a stochastic sampling algorithm.
We utilize the code {\sc Bilby}~\cite{BilbyInPrep}, which is a Python wrapper for carrying out parameter estimation with off-the-shelf samplers~\footnote{Examples of gravitational-wave injection and recovery with {\sc Bilby} can be found here: \url{https://git.ligo.org/lscsoft/bilby}}.
In this instance, we employ the nested sampling package {\sc PyMultiNest}~\cite{PyMultiNest2014} to sample the parameters describing gravitational-wave signals generated by waveform approximants available in {\sc LALSuite} \cite{LALSuite}.

\begin{figure}[t!]
	\centering
    \includegraphics[width=\linewidth]{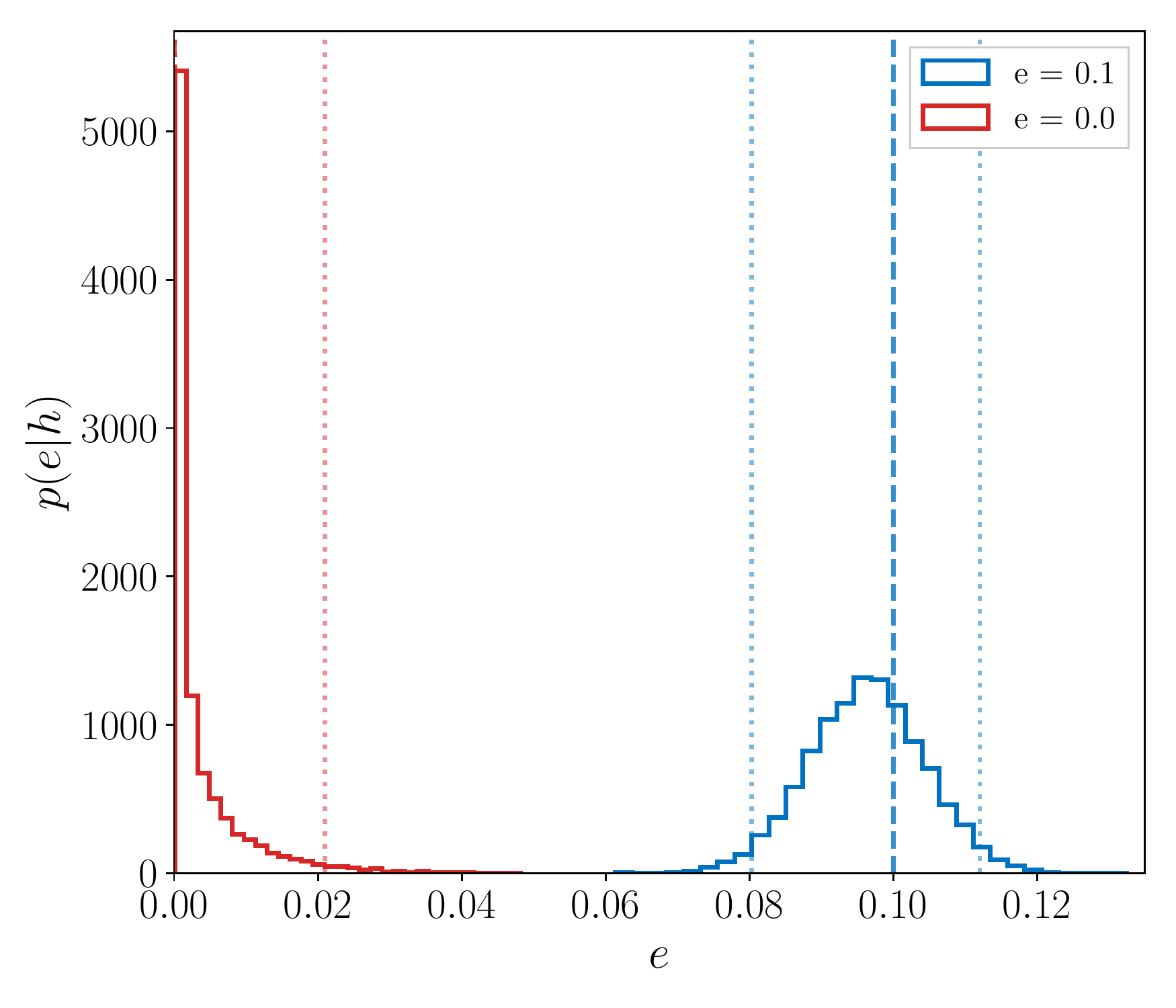}
    \caption{
    Posterior distributions of eccentricity at $\unit[10]{Hz}$ for two simulated GW150914-like events with eccentricities of $e = 0.1$ (blue) and $e = 0$ (red). The dotted vertical lines correspond to the 95\% credible intervals.
    }
    \label{fig:post_e}
\end{figure}

\begin{figure*}[t!]
	\centering
    \includegraphics[width=0.75\linewidth]{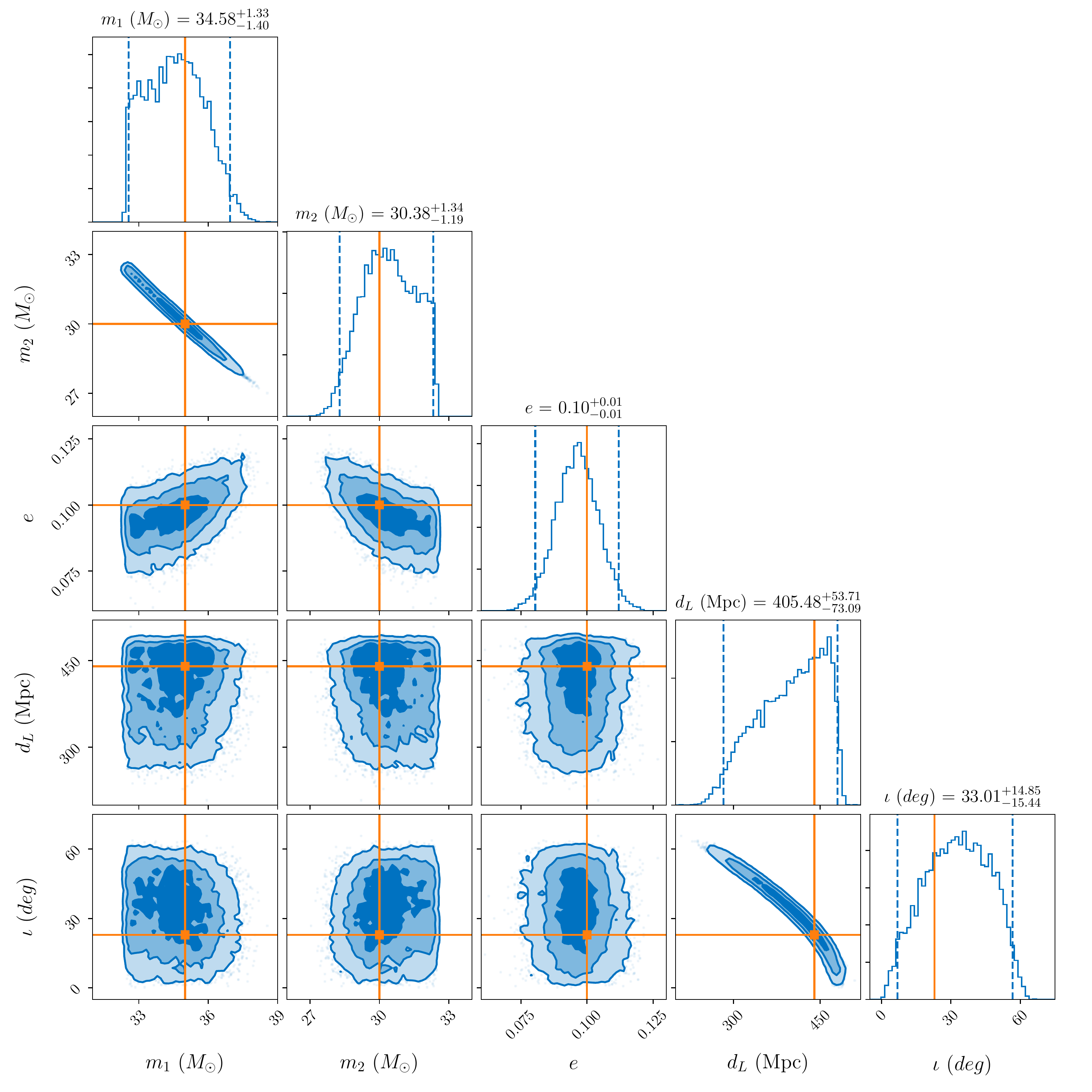}
    \caption{Posterior distributions for the primary black hole mass ($m_{1}$), secondary black hole mass ($m_{2}$), eccentricity ($e$), distance ($d_{L}$) and inclination ($\iota$). The waveform is an eccentric event ($e = 0.1$) with GW150914-like black hole masses and distance, with the true values given by the orange lines. Contours in the two-dimensional posteriors represent the 68\%, 95\% and 99\% confidence intervals. The mean recovered posteriors and 95\% confidence intervals are displayed at the top of each one-dimensional posterior distribution.}
    \label{fig:corner1}
\end{figure*}

While approximants that describe the full inspiral, merger and ringdown of eccentric, nonspinning BBH merger events exist~\cite{Hinder2017, Huerta2017a}, at the time of writing these models are yet to be implemented in {\sc LALSuite}.
In our analysis, we use a frequency-domain waveform approximant, {\sc EccentricFD} \cite{Huerta2014}, which models the $(l,|m|) = (2,2)$ mode of nonspinning, black hole binaries on precessing eccentric orbits.
It includes gravitational-wave phase corrections up to the 3.5 post-Newtonian (PN) order. It can be used for all black hole masses observable within the Advanced LIGO and Virgo frequency bands (see Figure 5 in~\cite{Huerta2014}), and has been shown to accurately reproduce an equivalent time-domain waveform ($< 3\%$ phase difference) in $\unit[12]{M_{\odot}}$ BBH with  eccentricities up to $e=0.4$ at $\unit[10]{Hz}$ (Throughout this paper, we measure eccentricity $e$ at $\unit[10]{Hz}$ unless otherwise stated.)
Preliminary studies of the expected degeneracy between eccentricity and spin corrections suggest that analyzing BBH mergers with noneccentric (quasicircular), spinning waveform models can result in the misclassification of eccentric events as coming from quasicircular BBH \cite{Huerta2018}, and may introduce potential biases in the recovered binary parameters \cite{Klein2018}.
In addition, recent work by \cite{Rebei18} using numerical relativity simulations of eccentric BBH has shown the inclusion of higher order modes can significantly affect the waveforms from eccentric BBH with large mass ratios.
We proceed with {\sc EccentricFD}, which is currently the only frequency-domain eccentric approximant available in {\sc LALSuite}, acknowledging its limitations and recommend that this analysis should be updated as more sophisticated approximants become available.

Using the current implementation of {\sc LALSuite}, we are unable to alter the input argument of periapsis in {\sc EccentricFD}.
While this is not ideal for full eccentric parameter estimation, we are more interested in the effect the magnitude of the eccentricity (the parameter e) has on the waveform. Hence, throughout this work the initial value of $\omega$ is always zero.
Future work should sample over $\omega$.

To avoid potential biases induced by the sharp cutoff at the end of the waveform \cite{Mandel2014} we introduce a frequency cut. This cut is set to be below the frequency at which the waveform terminates.
When the frequency cut is not included, the mass posterior exhibits multimodality that is not present when the cut is employed.
Since the binary circularizes rapidly, we do not expect the merger and ringdown to include a strong signature of the eccentricity.
However, by measuring the merger and ringdown, it is possible to better constrain other parameters, which may be covariant with eccentricity, thereby improving the measurement of eccentricity indirectly.
Future analyses that include merger and ringdown may therefore achieve better constraints on eccentricity.

\begin{table}[b]
   \caption{Source properties of the simulated event.}
   \begin{ruledtabular}
   \renewcommand*{\arraystretch}{1.1}
       \begin{tabular}{llll}
          Primary black hole mass & $m_{1}$ && $\unit[35]{M_{\odot}}$ \\
          Secondary black hole mass & $m_{2}$ && $\unit[30]{M_{\odot}}$  \\
          Eccentricity at $\unit[10]{Hz}$ & $e$ && 0.1 \\
          Luminosity distance & $d_{L}$ && $\unit[440]{Mpc}$ \\
          Inclination angle & $\iota$ && $\unit[22.9]{^\circ}$ \\
          Polarization angle & $\psi$ &&  $5.7^{\circ}$\\
          Phase at coalescence & $\phi$ && $68.7^{\circ}$ \\
          Right ascension & $\alpha$ && $\unit[3.7]{hrs}$ \\
          Declination & $\delta$ && $-31.7^{\circ}$ \\
          Network S/N & $\rho$ && 69.2\\
      \end{tabular}
  \end{ruledtabular}
  \label{table:params}
  \end{table}

As a demonstration of this formalism, we carry out parameter estimation on a simulated event injected into Gaussian noise colored to match the amplitude spectral densities of Advanced LIGO and Virgo at their design sensitivities. 
We begin by assuming that a BBH merger has been detected by some other algorithm~\cite{gstlal,pycbc}, either a dedicated search for compact binaries or an excess-power ``burst'' search~\cite{Tai2014,Coughlin2015,Tiwari2016}.
For a discussion of the impact of eccentricity on compact binary detection, see~\cite{Brown2010,Huerta2013}.
The luminosity distance and black hole masses are within the credible range of GW150914~\cite{150914improved}. 
A summary of the simulated binary parameters is provided in Table~\ref{table:params}.
We assume this event is observed by two design sensitivity Advanced LIGO detectors situated at the Hanford and Livingston sites, and the Virgo detector in Italy, with noise sensitivity curves from \cite{LIGOpsd} and \cite{Virgopsd} respectively. Each of the waveforms start at $f_{\min} = \unit[10]{Hz}$, the minimum frequency that can be observed within the Advanced LIGO band at design sensitivity.
We employ uniform prior distributions for the primary and secondary masses on the interval $(\unit[5]{M_\odot},\unit[60]{M_\odot})$ and a uniform-in-volume prior on the distance $\pi(d_L)\propto d_L^2$ from $\unit[100-2000]{Mpc}$.
We use standard priors for the extrinsic variables.
For eccentricity we use log-uniform priors (at $\unit[10]{Hz}$) on the interval $10^{-4} < e < 0.4$.
Since the {\sc EccentricFD} approximant does not accommodate spin, we set $\vec{S}_{1}=\vec{S}_{2}=0$.
Future analyses should allow for spin as more sophisticated approximants become available. 

In Figure \ref{fig:post_e}, we plot the eccentricity posterior probability distributions for two simulated BBH inspirals: one with $e = 0.1$ (blue) and the other with $e = 0$ (red).
The true eccentricity of the eccentric event, indicated by the dashed blue line, is within the 95\% credible interval of the distribution peak. The posterior for the quasicircular event rules out eccentricities greater than 0.02 with 95\% confidence.
We also plot the posterior probability distributions for the masses, eccentricity, distance and inclination for the eccentric simulation in Figure \ref{fig:corner1}. The contours in each two-dimensional posterior distribution represents the 68\%, 95\% and 99\% confidence intervals.
A figure showing the posterior distributions for the parameters not included in Figure \ref{fig:corner1} can be found in Appendix~\ref{appdx:posteriors}.

In order to determine the minimum distinguishable eccentricity for a BBH like GW150914, we employ both an (overly) optimistic ``overlap'' method and model selection.
The overlap method is useful for a quick back-of-the-envelope answer. However, we later show that it does not provide a reliable estimate when compared with a Bayes factor calculation, which includes covariances between the binary parameters.

\subsection{\label{subsec:overlap}Waveform overlap}

The level of mismatch between two gravitational waveforms can be quantified by calculating the overlap function, first used in~\cite{Flanagan1998}.
The overlap between an eccentric waveform $h_{\varepsilon}$ and a quasicircular waveform $h_{0}$, maximizing over the time and phase ($t_{0}, \phi_{0}$) of the quasicircular waveform, is
\begin{equation}
    \mathcal{O} = \max_{t_{0},\phi_{0}} \frac{\langle h_{0}|h_{\varepsilon}\rangle}{\sqrt{\langle h_{0}|h_{0}\rangle\langle h_{\varepsilon}|h_{\varepsilon}\rangle}},
	\label{eq:overlap}
\end{equation}
where
\begin{equation}
\left<a|b\right>\equiv4{\rm Re}\int_0^{\infty}df\frac{\tilde{a}(f)\tilde{b}^\star(f)}{S_h(f)},
\end{equation}
in which $S_h(f)$ is the noise power spectral density.
The overlap takes on values between -1 (corresponding to waveforms $180^{\circ}$ out of phase) and 1 (for identical waveforms).

Using the overlap rule of thumb, a gravitational waveform that includes eccentricity is distinguishable from the quasicircular waveform if
\begin{align}
1 - \mathcal{O}  \gtrsim \rho_{0}^{-2} ,
\end{align}
where $\rho_{0}^{2} = \langle h_{0}|h_{0} \rangle$ is the optimal matched-filter S/N of the quasicircular waveform.
The value of $1-\mathcal{O}$ is referred to as the ``mismatch", and is used for quantifying the required accuracy of waveform templates for detecting a gravitational-wave signal \cite{Flanagan1998, Lindblom2008}. It is inversely proportional to the S/N of the quasicircular waveform, allowing for smaller eccentricities to be probed in louder events. 

While the overlap is convenient for obtaining rough  estimates, it is not trustworthy due to its reliance on prior knowledge of the precise binary parameters.
As a result, the minimum distinguishable eccentricity we find via this method is overly optimistic.
It is possible to derive a generalized overlap reduction, which uses a $\chi^2$ factor to take into account covariance between parameters~\cite{Baird}.
However, even the generalized overlap is overly optimistic since it relies on a Fisher matrix approximation.

\subsection{\label{subsec:lnBF}Bayes factor}
In order to correctly take into account covariances between different parameters, we employ Bayesian model selection.
We calculate a Bayes factor comparing two hypotheses.

\begin{enumerate}[(i)]
       \item \textbf{Null hypothesis:} the signal is accurately described by a template in which the orbital eccentricity is zero. That is, the prior on eccentricity is a delta function, $\pi(e)=\delta(e)$, centered at zero.
    \item \textbf{Eccentric hypothesis:} the signal is best described by a waveform template in which eccentricity is nonzero. We employ a prior on eccentricity $\pi(e)$ that is log-uniform between $10^{-4}$ and $0.4$.
\end{enumerate}

In order to determine which hypothesis is favored by the data, we compute the Bayes factor
\begin{equation}\label{eqn:BF}
    \mathcal{B} = \frac{\mathcal{Z}_{\varepsilon}(h)}{\mathcal{Z}_{0}(h)} = 
    \frac{\int d\vec{\theta}\,de\,
    \mathcal{L}
(h|\vec{\theta},e)\pi(\vec{\theta})\pi(e)}
    {\int d\vec{\theta}\,\mathcal{L}
 (h|\vec{\theta},e=0)\pi(\vec{\theta})} .
\end{equation}
Here, $\pi(\vec{\theta})$ is the prior on binary parameters except for eccentricity and $\pi(e)$ is our log-uniform prior on eccentricity. 
The variable ${\cal L}(h|\vec\theta)$ is the likelihood.
Following convention, we employ a threshold of $|\ln(\mathcal{B})| > 8$ as the point at which one model becomes significantly preferred over the other.
The eccentricity for which $\ln(\mathcal{B}) > 8$ is the minimum detectable value.

\section{\label{sec:BFvsO}Distinguishing Eccentric Inspirals}

\begin{figure*}[t!]
    \centering
    \includegraphics[width=\linewidth]{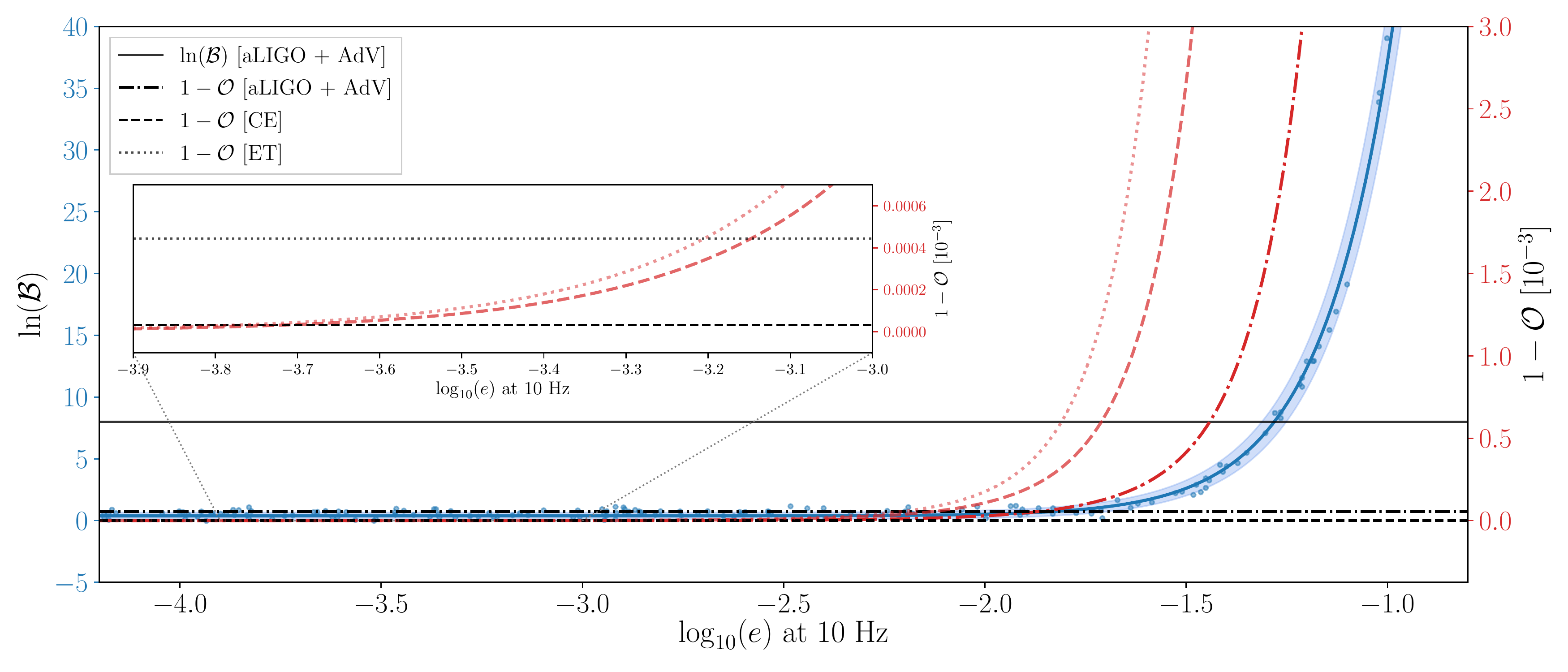}
    \caption{Overlap (red) for Advanced LIGO and Virgo (dash-dotted), CE (dashed) and ET (dotted), and the Bayes factor (blue points fit by nonlinear least squares, shaded region is the five-sigma fit error) as a function of the initial eccentricity, defined at $f_{\mathrm{GW}} = \unit[10]{Hz}$ for events with GW150914-like parameters. The horizontal lines correspond to thresholds of $\ln(\mathcal{B}) = 8$ and $1 - \mathcal{O} = 1/\rho_{0}^{2}$. The inset shows a zoom-in on the points at which the waveform overlap results for the CE and ET detectors cross their respective thresholds (dashed and dotted horizontal lines).}
    \label{fig:OverVsBF}
\end{figure*}

We compare the two methods (overlap and Bayes factor) using a set of simulated eccentric BBH inspiral events with component masses and distance similar to GW150914. The waveforms are generated using the parameters listed in Table~\ref{table:params}, with eccentricities ranging between $10^{-4}$ and $0.4$ at 10 Hz. 

Carrying out parameter estimation that takes advantage of the larger bandwidth of the two proposed third-generation detectors, Cosmic Explorer (CE) \cite{CosmicExplorer} and the Einstein Telescope (ET) \cite{EinsteinTelescope}($f_{\mathrm{min, CE}} = \unit[5]{Hz}$, $ f_{\mathrm{min, ET}} = \unit[1]{Hz}$), is computationally difficult because binary waveforms quickly become longer as the minimum frequency is reduced. 
The computational challenge is compounded by the fact that the likelihood function becomes highly peaked for events observed in third-generation detectors, and so the evidence calculation takes longer to converge.

\begin{table}[b]
   \caption{
   Minimum distinguishable eccentricities produced via the Bayesian method for Advanced LIGO/Virgo and the waveform overlap method for Advanced LIGO/Virgo, CE and ET.
   We assume an event similar to GW150914.
   }
   \begin{ruledtabular}
   \renewcommand*{\arraystretch}{1.1}
       \begin{tabular}{cccc}
          Method & Detectors & $f_{\min}$ (Hz) & $e_{\mathrm{min}}$  \\ \hline \rule{0pt}{2.6ex}
          Bayesian & aLIGO + AdV  & 10 & $0.052$ \\
          Overlap  & aLIGO + AdV  & 10 & $0.014$ \\
          Overlap  & CE $\times$ 2   & 5  & $0.00017$ \\
          Overlap  & CE $\times$ 2   & 10 & $0.00019$ \\
          Overlap  & ET $\times$ 2   & 1  & $0.00024$ \\
          Overlap  & ET $\times$ 2   & 10 & $0.00062$
      \end{tabular}
  \end{ruledtabular}
  \label{table:emin}
  \end{table}

For these reasons, we do not apply our Bayesian method using the sensitivity curves of CE and ET. 
However, we can use the waveform overlap method to place optimistic upper limits on the minimum distinguishable eccentricity that can be observed with these detectors.
We calculate the overlap two ways: using $f_\text{min}=\unit[10]{Hz}$ for comparison with LIGO and Virgo, and using a lower $f_\text{min}$, $\unit[5]{Hz}$ for CE and $\unit[1]{Hz}$ for ET.
For Advanced LIGO and Virgo, we assume the same detector network used in Section~\ref{sec:formalism} for their respective overlap and Bayesian analyses. When applying the overlap method to the third generation detectors, we assume a network of either two CE or ET detectors located at the Hanford and Livingston sites with sensitivity curves from \cite{CEpsd} and \cite{ETpsd, Hild2011}.
Detector networks with additional detectors will be able to probe lower eccentricities.

In Figure \ref{fig:OverVsBF}, we plot  the overlap and the Bayes factor as a function of eccentricity.
The thresholds for detectability are indicated with horizontal lines.
The associated minimum detectable eccentricities ($e_\text{min}$) are given in Table \ref{table:emin}.
For Advanced LIGO and Virgo, the overlap method yields a minimum distinguishable eccentricity of $e_\text{min}=0.014$ while the Bayes factor technique yields $e_\text{min}=0.052$.
The fact that the Bayes factor technique yields a significantly larger minimum detectable eccentricity highlights the limits of the overlap method, which does not include covariance between different binary parameters.
For a comparison of this result with~\cite{Gondan2018}, refer to Appendix~\ref{appdx:Gondan}.

As we discuss below, the $\unit[10]{Hz}$ eccentricity of globular cluster triples is likely to be well above this level.
Triples are likely to constitute $\approx5\%$ of the mergers in globular clusters.
Thus, if globular clusters are the primary source of BBH mergers, it should be possible for advanced detectors to infer this with ${\cal O}(100)$ events.
  
Using the overlap method, and setting $f_\text{min}=\unit[10]{Hz}$, the minimum eccentricity for third-generation detectors are $e_\text{min}=1.9 \times 10^{-4}$ for CE and $e_\text{min}=6.2 \times 10^{-4}$ for ET.
This represents an improvement over the minimum distinguishable eccentricity observable by Advanced LIGO and Virgo for GW150914-like events by almost two-orders of magnitude.
Repeating the calculation with smaller values of $f_\text{min}$, we obtain $e_\text{min}=1.7 \times 10^{-4}$ for CE integrating from $\unit[5]{Hz}$, and $e_\text{min}=2.4 \times 10^{-4}$ for ET integrating from $\unit[1]{Hz}$.
Note that the eccentricity is still referenced to $\unit[10]{Hz}$ no matter the minimum observing frequency.
Additional details, exploring how the Bayes factor scales with both mass and matched-filter signal-to-noise ratio are explored in Appendix~\ref{appdx:BFscaling}.

\section{\label{sec:populations}Eccentric Populations}

\begin{figure}[t!]
   \centering
   \includegraphics[width=\linewidth]{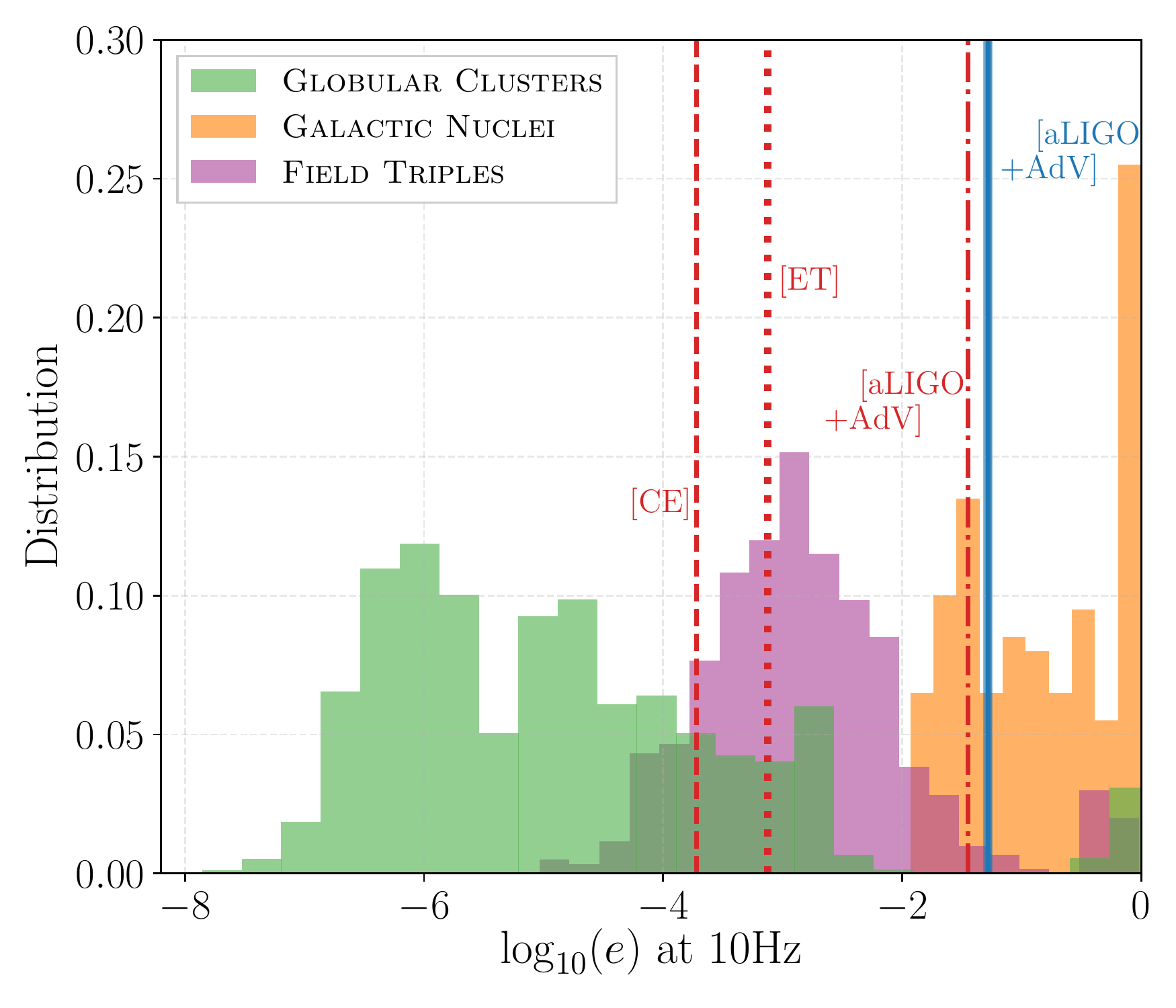}
    \caption{Eccentricity distributions at $f_{\mathrm{GW}} = \unit[10]{Hz}$ for different eccentric BBH with GW150914-like masses.
    The green distribution shows binaries formed in globular clusters \cite{Rodriguez2018a}.
    From left to right: the first peak is from ejected binaries, the second is from two-body mergers in the binary, and the third peak is from three-body mergers~\cite{Samsing2018a}.
    Field triples are shown in purple \cite{Antonini2017}.
    Direct capture within galactic nuclei is shown in orange \cite{Oleary2009}.
    Vertical lines correspond to the different minimum distinguishable eccentricities calculated from the overlap (red) and Bayes factor (blue) methods for the eccentric GW150914-like events analyzed in Table \ref{table:emin}.}
    \label{fig:eccDist}
\end{figure}

While we have drawn attention to the recent predictions of high eccentricity from three-body mergers in globular clusters~\cite{Rodriguez2018a, Samsing2018a}, there are a number of other predicted origins for eccentric mergers.
Here, we compare the eccentricity distributions (shown in Figure~\ref{fig:eccDist}) from three models of eccentric BBH formation to the minimum distinguishable eccentricities found in Section \ref{sec:BFvsO}.
The three models that we consider are:
\begin{enumerate}[(i)]
    \item {\sc Globular Clusters} (green distribution in Figure~\ref{fig:eccDist}).
    This is our fiducial model from~\cite{Rodriguez2018a}---see also~\cite{Samsing2018a}---which includes contributions from ejected mergers (first peak, $\sim\unit[10^{-5}-10^{-3}]{Hz}$), two-body mergers in the globular cluster (second peak, $\sim\unit[10^{-4}-10^{-2}]{Hz}$), and three-body mergers (third peak, $\sim\unit[1-100]{Hz}$).
The merger rate from globular clusters is  uncertain.
    \item {\sc Galactic Nuclei} (orange distribution in Figure~\ref{fig:eccDist}).
    This model, based on~\cite{Oleary2009}, posits that binary black holes merge dynamically in the dense stellar environment of a galactic nuclei.
    These environments are significantly more challenging to model than globular clusters, and so the eccentricity distribution is less certain than the globular cluster model.
Preliminary estimates of the merger rate for close flybys between BH in galactic nuclei are $\unit[\sim0.02]{yr^{-1}Gpc^{-3}}$ for $10$ M$_{\odot}$~\cite{Tsang2013} black holes.
This is comparatively low given the observed total BBH merger rate of $\unit[12-213]{yr^{-1}Gpc^{-3}}$~\cite{Abbott170104}.
Recent work has investigated whether eccentric binaries are formed near supermassive black holes through the Kozai-Lidov mechanism, which may have merger rates more in-line with the observed rate~\cite{Antonini2012,Hoang2018,Hamers2018}.

    \item {\sc Field Triples} (purple distribution in Figure~\ref{fig:eccDist}).
    This model invokes hierarchical black hole triples undergoing Lidov-Kozai oscillations, which form as the result of isolated field triple evolution~\cite{Antonini2017}.
	Unlike the other two models, these mergers are not dynamical.
    The eccentricity distribution for hierarchical triples presented in \cite{Antonini2017} was derived from the output of complex three-body simulations.
    We approximate the eccentricity distribution of the hierarchical triple systems as a Gaussian in $\log_{10}(e)$ with a mean of $\mu = -3$ and a variance of $\sigma = 0.7$.
    A small fraction of mergers originating from BH triple systems ($\sim 5\%$) are predicted to enter the Advanced LIGO band with extreme eccentricities of nearly unity~\cite{Antonini2017}.
    Preliminary estimates of the merger rate for field triples range from $\unit[0.14 - 6]{yr^{-1}Gpc^{-3}}$ \cite{Silsbee2017, Antonini2017}, which is small compared to the total observed merger rate.
    However, the rate of eccentric field triples may be comparable to the rate from globular clusters if natal kicks are small~\cite{Silsbee2017,Antonini2017}. The field triple rate may also be increased in low-metallicity environments~\cite{Rodriguez2018b}.
\end{enumerate}

For the {\sc Galactic Nuclei} model, it is necessary to evolve the eccentricity distribution at formation to the LIGO band, with the initial semi-major axis and eccentricities calculated from the analytic methods outlined in \cite{Oleary2009, Kocsis2012}.
In order to evolve the system forward in time, we use the analytic expression describing the evolution of semi-major axis as a function of eccentricity from \cite{Peters1964},
\begin{equation}\label{eqn:analytic}
    a(e) = \frac{c_{0}e^{\small{12/19}}}{(1-e^{2})}\Big[1 + \frac{121}{304}e^{2}\Big]^{\small{870/2299}},
\end{equation}
where $c_{0}$ is a constant, the value of which depends on the initial semi-major axis and eccentricity as
\begin{equation}
    c_{0} = \frac{a_{0}(1 - e_{0}^{2})}{e^{12/19}_{0} \Big[1 + \frac{121}{304}e_{0}^{2}\Big]^{870/2299}}.
\end{equation}
The frequency of the emitted gravitational waves evolves according to~\cite{Wen2003}
\begin{equation}\label{eqn:eccfreq}
	f_\text{gw}(a, e) \simeq \frac{\sqrt{G M_{tot}}}{\pi} \frac{(1+e)^{1.1954}}{[a(1-e^{2})]^{3/2}}.
\end{equation}
We evolve the semi-major axis and eccentricity of the {\sc Galactic Nuclei} BBH until the peak gravitational-wave frequency enters the Advanced LIGO and Virgo band at $f_\text{min}=\unit[10]{Hz}$. 

The eccentricity distributions for these three formation models are presented in Figure \ref{fig:eccDist}, with the minimum distinguishable eccentricities from Table \ref{table:emin} for Advanced LIGO/Virgo, CE and ET represented by the vertical lines.
The highly eccentric peak in the {\sc Globular Cluster} distribution results from the 5\% of mergers originating from three-body driven mergers.
These three-body mergers enter the advanced-detector band with sufficient eccentricity that it is likely they can be distinguished from circular binaries.
This conclusion is robust if one makes slightly different assumptions about the velocity dispersion relation and/or black hole density in globular clusters.
If globular clusters are the dominant source of binary black hole mergers, we will probably know after approximately $20$ detections.
If there is no evidence of eccentricity after $100$ mergers, it may be possible to conclude that globular clusters play a subdominant role in creating black hole binary mergers.

We can also conclude from Figure~\ref{fig:eccDist} that more than half of events in the {\sc Galactic Nuclei} model can be distinguished from circular binaries.
However, this result is probably best taken with a grain of salt since it is difficult to model the dynamics of galactic nuclei.
We also see that it is possible to measure eccentricity in a few percent of events from the {\sc Field Triples} model.
It may be possible to distinguish between the {\sc Field Triples} model and {\sc Globular Clusters} model by looking for binaries with eccentricities of $e\approx 10^{-1.5}$, which are only present in the {\sc Field Triples} model.

Turning our attention to third-generation detectors, we see that both CE and ET may be capable of catching the second peak in the {\sc Globular Cluster} model, which is due to two-body BBH mergers within the globular cluster.
Note the CE and ET sensitivities should be taken with some level of caution since they are derived using the optimistic overlap calculation.
Measuring two different components of the {\sc Globular Cluster} distribution could provide a powerful confirmation of the globular cluster paradigm.

\section{\label{sec:discussion}Discussion}

We do not know the precise formation mechanism of the BBH detected by Advanced LIGO and Virgo.
The capability of detecting eccentricity in the orbits of BBH systems would help us to understand BBH formation and allow us to probe the environment in which these systems reside.
In this paper, we demonstrate Bayesian parameter estimation using the inference code {\sc Bilby}~\cite{BilbyInPrep} with the {\sc PyMultiNest} sampler.
We calculate the sensitivity of Advanced LIGO and Virgo to eccentricity using Bayesian model comparison, which is contrasted against a naive sensitivity calculated using an overlap factor.
For an event with similar masses and distance to GW150914, the minimum distinguishable eccentricity--determined using Bayesian model selection--is $e=0.052$ at $\unit[10]{Hz}$.
All else equal, we find it is easier to detect the eccentricity of relatively lower mass systems (see Appendix~\ref{appdx:BFscaling}).

We then compare the minimum detectable eccentricity to distributions for three different models: {\sc Globular Clusters}, {\sc Galactic Nuclei}, and {\sc Field Triples}.
From this comparison, we find that second-generation detectors should be able to find evidence for or against the hypothesis that the observed BBH merger rate is dominated by globular cluster binaries with about 20-100 events.
The globular cluster hypothesis will gain support if $\approx5\%$ of Advanced LIGO/Virgo-band binaries exhibit large eccentricities, which can result from three-body driven mergers.
This result is relatively robust to different assumptions about the velocity dispersion and black hole density in globular clusters.
Third-generation detectors may be able to observe two-body mergers with much lower eccentricity, which would further cement the globular cluster paradigm.

We expect future studies, which utilize more complete eccentric waveform models that include the merger and ringdown phases, will provide a more realistic picture of how measurable eccentricity is in realistic BBH merger events. In addition, since the {\sc{EccentricFD}} model is limited to nonspinning BBH, we are unable to explore the effect of potential degeneracies between spin-orbit coupling and eccentricity in our measurements. Hence the analysis presented in this paper will be updated in the future as more sophisticated approximants become available.

The framework described in this paper can also be utilized in population studies, where an ensemble of eccentric detections could allow advanced detectors to probe smaller eccentricities than we report. Implementation of this is left for future studies.

\section*{Acknowledgements}
We would like to thank Johan Samsing and Yuri Levin for their insightful conversations. We also give thanks to Carl Johan-Haster, Ed Porter, John Veitch, Thomas Dent, Harald Pfeiffer and Simon Stevenson for their valuable comments and corrections. We thank the referee for their suggestions.
This work is supported through Australian Research Council (ARC) Centre of Excellence CE170100004. 
M.E.L. receives support from the Australian Government Research Training Program and ARC Laureate Fellowship FL150100148. 
E.H.T. is supported through ARC Future Fellowship FT150100281.
P.D.L. is supported through ARC Future Fellowship FT160100112 and ARC Discovery Project DP180103155.
This work made use of the OzSTAR national HPC facility.
This paper has been assigned the document ID LIGO-P1800138.

\begin{appendix}
\section{Additional Posterior Distributions}\label{appdx:posteriors}

\begin{figure*}[h]
	\centering
    \includegraphics[width=0.75\linewidth]{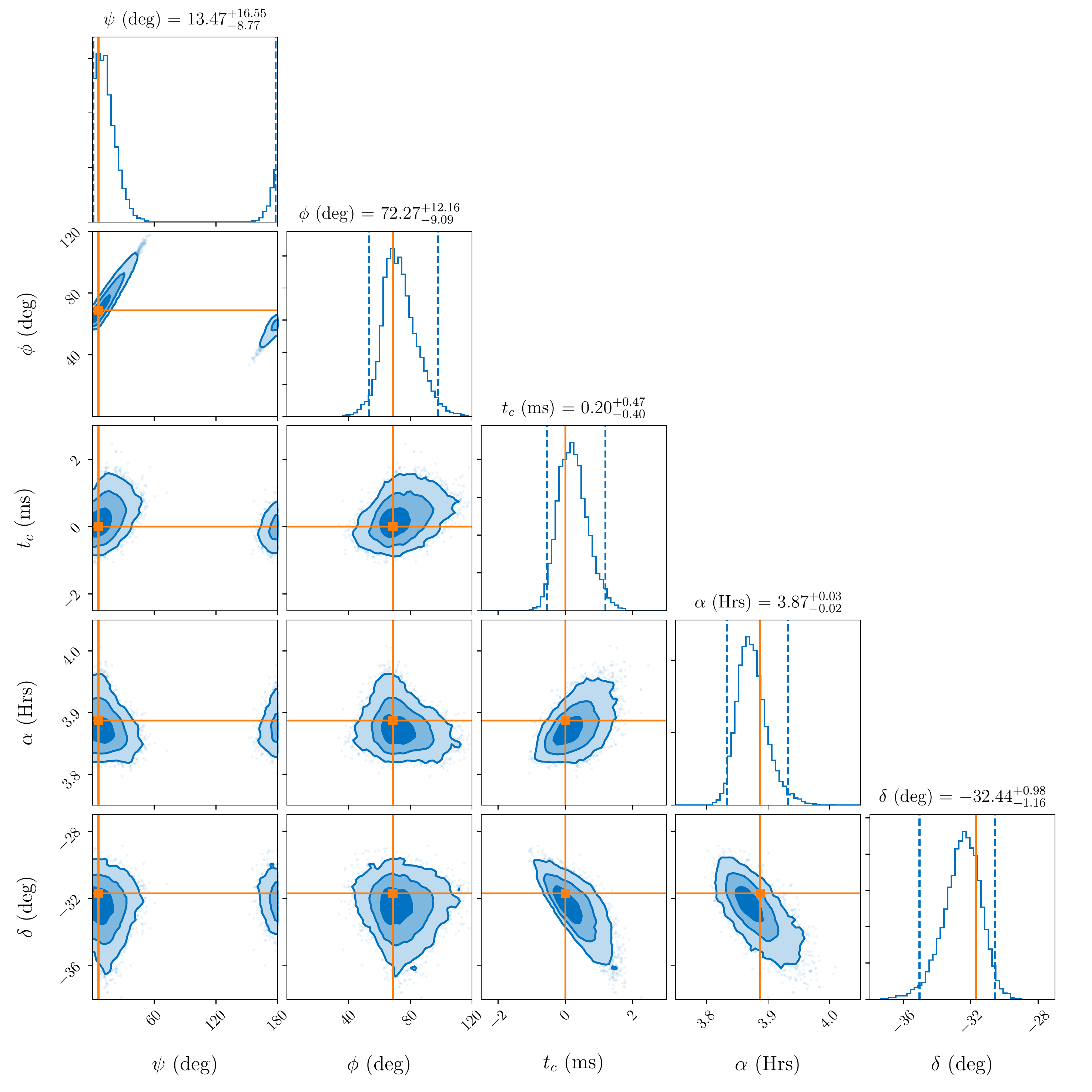}
    \caption{Posterior distributions for the select waveform parameters not shown in Figure \ref{fig:corner1}. Contours in the two-dimensional posteriors represent the 68\%, 95\% and 99\% confidence intervals and the true values are indicated by orange lines. Note the coalescence time ($t_{c}$) is in units of ms either side of the true value of $\unit[1180002601]{s}$.}
    \label{fig:corner2}
\end{figure*}

In Figure~\ref{fig:corner2} we present posterior distributions for the gravitational-wave polarization angle ($\psi$), binary phase at coalescence ($\phi$), coalescence time ($t_{c}$), and the source location on the sky in right-ascension and declination ($\alpha$, $\delta$).

\section{Comparison with Fisher Matrix Results}\label{appdx:Gondan}

Here we compare the minimum eccentricity from our Bayes factor calculation to the eccentricity uncertainty derived in~\cite{Gondan2018} using a Fisher matrix calculation (see their Figure 4).
There are significant caveats that we must first make owing to the very different nature of these two calculations.
First, while both analyses consider events similar to GW150914, \cite{Gondan2018} considers a binary with an initial eccentricity of $e_0=0.9$ at formation, while we consider waveforms with $e\lesssim 0.4$ as they enter the Advanced detector frequency band at $f_\text{GW} = \unit[10]{Hz}$.
Given a reliable approximant, such highly eccentric waveforms are easier to detect than a less eccentric waveforms, however we are limited to the more modest eccentricities allowed by {\sc EccentricFD}.
Second, as we note above, the Fisher matrix calculation provides an optimistic result by modeling the likelihood function as a multivariate Gaussian, which it is not.
Third, \cite{Gondan2018} employs a new waveform model, which is not currently available for Bayesian parameter estimation using {\sc LALSuite}. This model has some similarities to {\sc EccentricFD} in that it models the $(l,|m|) = (2,2)$ mode of nonspinning, inspiral only, precessing, eccentric BBH. However it is restricted to leading order PN corrections to the gravitational-wave phase, while {\sc EccentricFD} includes corrections up to the 3.5PN order.
Fourth, we assume a different detector network and different sky locations.

Noting all of these caveats, we estimate the median uncertainty from Figure \ref{fig:eccDist} in~\cite{Gondan2018} to be $\approx 10^{-3.2}$.
Taking account the ratio of the different injection distances ($\unit[440]{Mpc}/\unit[100]{Mpc}$=4.4), we estimate the~\cite{Gondan2018} uncertainty to be $\sigma_e\approx0.003$ at $\unit[410]{Mpc}$.
It is impossible to directly compare the frequentist $\sigma_e$ to our Bayes factor, but speaking roughly, a log Bayes factor of eight is, in some sense, comparable to a five sigma detection.
Thus, accounting for the differences in distance, and accounting for the difference in sigmas, we estimate the~\cite{Gondan2018} uncertainty to be $5\sigma_e\approx0.014$ at $\unit[410]{Mpc}$.
Remarkably, this result is consistent with our waveform overlap result, but is $3.7$ times smaller than our value of $e_\text{min}=0.052$. This last point is to be expected given the apple-to-orange nature of this comparison.

\section{Scaling Relations}\label{appdx:BFscaling}

In this section, we discuss scaling relations for how $e_{\mathrm{min}}$ depends on the S/N and total system mass $M_{\mathrm{tot}}$.
Higher S/N yields more sensitive measurements of all parameters, including $e$.
The mass of the black holes determines the time taken for a binary to merge. The longer the binary spends in band, the easier it is to measure the effect of eccentricity.
We explore scaling relations in two ways:
\begin{enumerate}[(i)]
\item We vary the S/N of a set of eccentric events each with the same fixed binary parameters (including mass). 
The S/N is varied by adjusting the distance.
\item We inject events with total black hole masses of either $M_{\mathrm{tot}} = \unit[30]{M_{\odot}}$,  $\unit[60]{M_{\odot}}$ or $\unit[90]{M_{\odot}}$, with fixed S/N.
\end{enumerate}

While much of the S/N from $M_{\rm{tot}} = \unit[60]{M_{\odot}}$ and $\unit[90]{M_{\odot}}$ BBH mergers comes from the merger and ringdown, numerical relativity simulations suggest eccentric BBH circularize by the late insprial stage \cite{Hinder2008}. Hence the lack of merger and ringdown phases in {\sc EccentricFD} may not significantly affect our ability to measure eccentricity. Additional investigation with improved approximants will determine if this is true.

\begin{figure}[t!]
	\centering    
    \includegraphics[width=\linewidth]{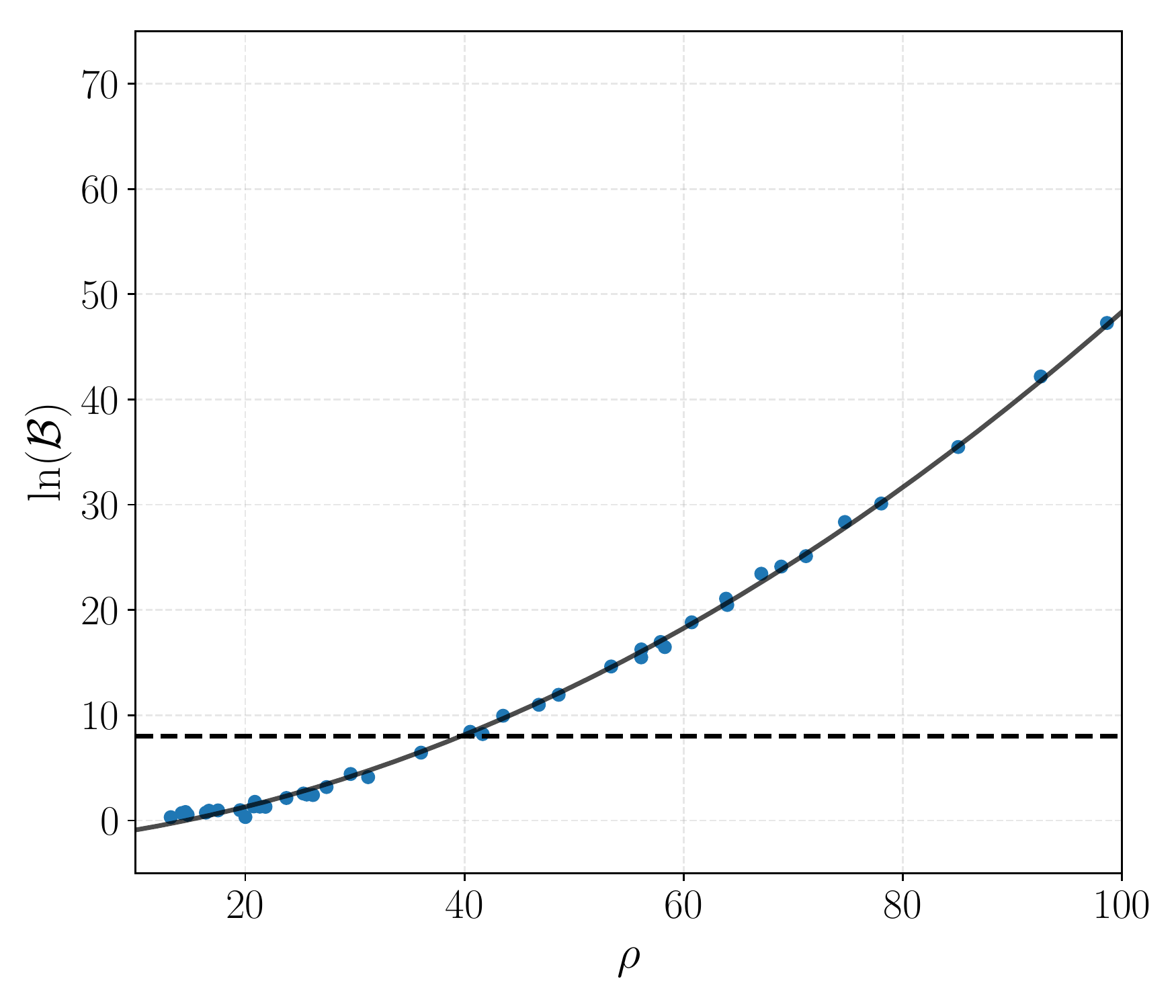}
    \caption{Growth of Bayes factor for events with increasing combined Hanford and Livingston S/N. The dashed horizontal line corresponds to the detection threshold at $\ln(\mathcal{B}) = 8$. The black curve corresponds to the line of best fit.}
    \label{fig:snrVsBF}
\end{figure}

\begin{figure}[t!]
	\centering    
    \includegraphics[width=\linewidth]{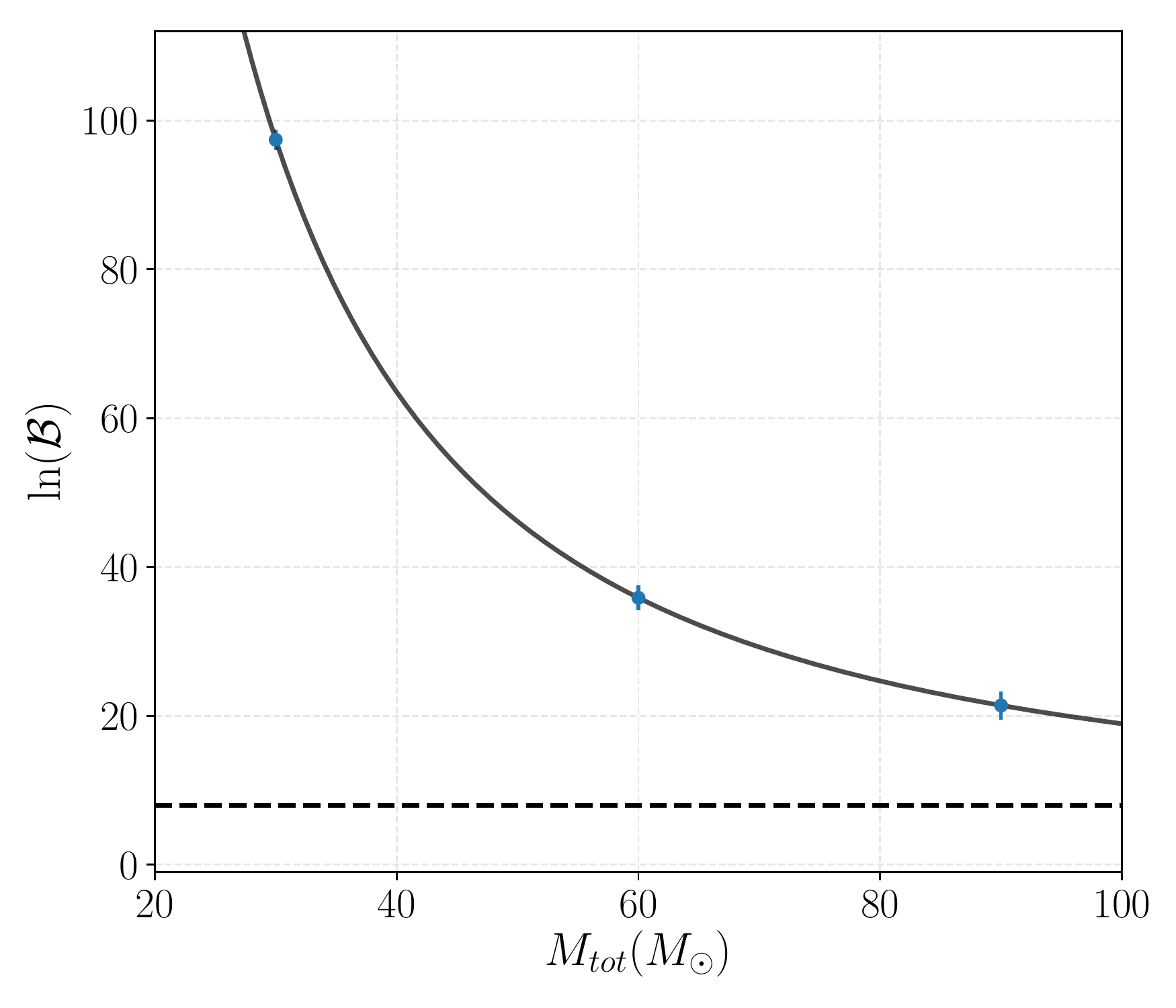}
    \caption{Comparison between the Bayes factor for events with total masses of $M_{\mathrm{tot}} = \unit[30]{M_{\odot}}$,  $\unit[60]{M_{\odot}}$ or $\unit[90]{M_{\odot}}$. Each of the events being compared has the same S/N and mass ratio. The black curve corresponds to the line of best fit. The dashed horizontal line corresponds to the detection threshold of $\ln(\mathcal{B}) = 8$.}
    \label{fig:MtotVsBF}
\end{figure}

In Figure~\ref{fig:snrVsBF}, we plot the eccentric-vs-circular log Bayes factor, $\ln({\cal B})$, as a function of the matched filter signal-to-noise ratio $\rho$.
The black curve shows a 2-degree polynomial fit.
In Fig~\ref{fig:MtotVsBF}, we plot $\ln({\cal B})$ as a function of mass given a fixed eccentricity of $e=0.075$ at $\unit[10]{Hz}$ and fixed S/N.
We observe that, all else equal, lower-mass systems provide more sensitive measurements of eccentricity than higher-mass systems.
This is likely because lower-mass systems have more cycles in the observing band.
Given a fixed signal-to-noise ratio, we expect mass to play the most important role (out of all the waveform parameters) in determining the detectability of eccentricity.

\end{appendix}

\bibliographystyle{apsrev4-1}
\bibliography{main}

\begin{thebibliography}{90}%
\makeatletter
\providecommand \@ifxundefined [1]{%
 \@ifx{#1\undefined}
}%
\providecommand \@ifnum [1]{%
 \ifnum #1\expandafter \@firstoftwo
 \else \expandafter \@secondoftwo
 \fi
}%
\providecommand \@ifx [1]{%
 \ifx #1\expandafter \@firstoftwo
 \else \expandafter \@secondoftwo
 \fi
}%
\providecommand \natexlab [1]{#1}%
\providecommand \enquote  [1]{``#1''}%
\providecommand \bibnamefont  [1]{#1}%
\providecommand \bibfnamefont [1]{#1}%
\providecommand \citenamefont [1]{#1}%
\providecommand \href@noop [0]{\@secondoftwo}%
\providecommand \href [0]{\begingroup \@sanitize@url \@href}%
\providecommand \@href[1]{\@@startlink{#1}\@@href}%
\providecommand \@@href[1]{\endgroup#1\@@endlink}%
\providecommand \@sanitize@url [0]{\catcode `\\12\catcode `\$12\catcode
  `\&12\catcode `\#12\catcode `\^12\catcode `\_12\catcode `\%12\relax}%
\providecommand \@@startlink[1]{}%
\providecommand \@@endlink[0]{}%
\providecommand \url  [0]{\begingroup\@sanitize@url \@url }%
\providecommand \@url [1]{\endgroup\@href {#1}{\urlprefix }}%
\providecommand \urlprefix  [0]{URL }%
\providecommand \Eprint [0]{\href }%
\providecommand \doibase [0]{http://dx.doi.org/}%
\providecommand \selectlanguage [0]{\@gobble}%
\providecommand \bibinfo  [0]{\@secondoftwo}%
\providecommand \bibfield  [0]{\@secondoftwo}%
\providecommand \translation [1]{[#1]}%
\providecommand \BibitemOpen [0]{}%
\providecommand \bibitemStop [0]{}%
\providecommand \bibitemNoStop [0]{.\EOS\space}%
\providecommand \EOS [0]{\spacefactor3000\relax}%
\providecommand \BibitemShut  [1]{\csname bibitem#1\endcsname}%
\let\auto@bib@innerbib\@empty
\bibitem [{\citenamefont {Aasi}\ \emph {et~al.}(2015)\citenamefont {Aasi} \emph
  {et~al.}}]{Aasi2015}%
  \BibitemOpen
  \bibfield  {author} {\bibinfo {author} {\bibfnamefont {J.}~\bibnamefont
  {Aasi}} \emph {et~al.},\ }\href {\doibase 10.1088/0264-9381/32/7/074001}
  {\bibfield  {journal} {\bibinfo  {journal} {Classical Quantum Gravity}\
  }\textbf {\bibinfo {volume} {32}},\ \bibinfo {pages} {074001} (\bibinfo
  {year} {2015})}\BibitemShut {NoStop}%
\bibitem [{\citenamefont {Acernese}\ \emph {et~al.}(2015)\citenamefont
  {Acernese} \emph {et~al.}}]{Acernese2015}%
  \BibitemOpen
  \bibfield  {author} {\bibinfo {author} {\bibfnamefont {F.}~\bibnamefont
  {Acernese}} \emph {et~al.},\ }\href {\doibase 10.1088/0264-9381/32/2/024001}
  {\bibfield  {journal} {\bibinfo  {journal} {Classical Quantum Gravity}\
  }\textbf {\bibinfo {volume} {32}},\ \bibinfo {pages} {024001} (\bibinfo
  {year} {2015})}\BibitemShut {NoStop}%
\bibitem [{\citenamefont {Abbott}\ \emph
  {et~al.}(2016{\natexlab{a}})\citenamefont {Abbott} \emph
  {et~al.}}]{Abbott150914}%
  \BibitemOpen
  \bibfield  {author} {\bibinfo {author} {\bibfnamefont {B.~P.}\ \bibnamefont
  {Abbott}} \emph {et~al.},\ }\href {\doibase 10.1103/PhysRevLett.116.061102}
  {\bibfield  {journal} {\bibinfo  {journal} {Phys. Rev. Lett.}\ }\textbf
  {\bibinfo {volume} {116}},\ \bibinfo {pages} {061102} (\bibinfo {year}
  {2016}{\natexlab{a}})}\BibitemShut {NoStop}%
\bibitem [{\citenamefont {Abbott}\ \emph
  {et~al.}(2016{\natexlab{b}})\citenamefont {Abbott} \emph
  {et~al.}}]{Abbott151226}%
  \BibitemOpen
  \bibfield  {author} {\bibinfo {author} {\bibfnamefont {B.~P.}\ \bibnamefont
  {Abbott}} \emph {et~al.},\ }\href {\doibase 10.1103/PhysRevLett.116.241103}
  {\bibfield  {journal} {\bibinfo  {journal} {Phys. Rev. Lett.}\ }\textbf
  {\bibinfo {volume} {116}},\ \bibinfo {pages} {241103} (\bibinfo {year}
  {2016}{\natexlab{b}})}\BibitemShut {NoStop}%
\bibitem [{\citenamefont {Abbott}\ \emph {et~al.}(2017)\citenamefont {Abbott}
  \emph {et~al.}}]{Abbott170104}%
  \BibitemOpen
  \bibfield  {author} {\bibinfo {author} {\bibfnamefont {B.~P.}\ \bibnamefont
  {Abbott}} \emph {et~al.},\ }\href {\doibase 10.1103/PhysRevLett.118.221101}
  {\bibfield  {journal} {\bibinfo  {journal} {Phys. Rev. Lett.}\ }\textbf
  {\bibinfo {volume} {118}},\ \bibinfo {pages} {221101} (\bibinfo {year}
  {2017})}\BibitemShut {NoStop}%
\bibitem [{\citenamefont {{Abbott}}\ \emph {et~al.}(2017)\citenamefont
  {{Abbott}} \emph {et~al.}}]{Abbott170608}%
  \BibitemOpen
  \bibfield  {author} {\bibinfo {author} {\bibfnamefont {B.~P.}\ \bibnamefont
  {{Abbott}}} \emph {et~al.},\ }\href {\doibase 10.3847/2041-8213/aa9f0c}
  {\bibfield  {journal} {\bibinfo  {journal} {Astrophys. J. Lett.}\ }\textbf
  {\bibinfo {volume} {35}},\ \bibinfo {pages} {L35} (\bibinfo {year}
  {2017})}\BibitemShut {NoStop}%
\bibitem [{\citenamefont {Abbott}\ \emph {et~al.}(2017)\citenamefont {Abbott}
  \emph {et~al.}}]{Abbott170814}%
  \BibitemOpen
  \bibfield  {author} {\bibinfo {author} {\bibfnamefont {B.~P.}\ \bibnamefont
  {Abbott}} \emph {et~al.},\ }\href {\doibase 10.1103/PhysRevLett.119.141101}
  {\bibfield  {journal} {\bibinfo  {journal} {Phys. Rev. Lett.}\ }\textbf
  {\bibinfo {volume} {119}},\ \bibinfo {pages} {141101} (\bibinfo {year}
  {2017})}\BibitemShut {NoStop}%
\bibitem [{\citenamefont {Abbott}\ \emph
  {et~al.}(2016{\natexlab{c}})\citenamefont {Abbott} \emph
  {et~al.}}]{AbbottO1Cat}%
  \BibitemOpen
  \bibfield  {author} {\bibinfo {author} {\bibfnamefont {B.~P.}\ \bibnamefont
  {Abbott}} \emph {et~al.},\ }\href {\doibase 10.1103/PhysRevX.6.041015}
  {\bibfield  {journal} {\bibinfo  {journal} {Phys. Rev. X}\ }\textbf {\bibinfo
  {volume} {6}},\ \bibinfo {pages} {041015} (\bibinfo {year}
  {2016}{\natexlab{c}})}\BibitemShut {NoStop}%
\bibitem [{\citenamefont {Belczynski}\ \emph
  {et~al.}(2016{\natexlab{a}})\citenamefont {Belczynski}, \citenamefont {Holz},
  \citenamefont {Bulik},\ and\ \citenamefont {O'Shaughnessy}}]{Belczynski2016}%
  \BibitemOpen
  \bibfield  {author} {\bibinfo {author} {\bibfnamefont {K.}~\bibnamefont
  {Belczynski}}, \bibinfo {author} {\bibfnamefont {D.~E.}\ \bibnamefont
  {Holz}}, \bibinfo {author} {\bibfnamefont {T.}~\bibnamefont {Bulik}}, \ and\
  \bibinfo {author} {\bibfnamefont {R.}~\bibnamefont {O'Shaughnessy}},\ }\href
  {\doibase 10.1038/nature18322} {\bibfield  {journal} {\bibinfo  {journal}
  {Nature (London)}\ }\textbf {\bibinfo {volume} {534}},\ \bibinfo {pages}
  {512} (\bibinfo {year} {2016}{\natexlab{a}})}\BibitemShut {NoStop}%
\bibitem [{\citenamefont {Heger}\ \emph {et~al.}(2003)\citenamefont {Heger},
  \citenamefont {Fryer}, \citenamefont {Woosley}, \citenamefont {Langer},\ and\
  \citenamefont {Hartmann}}]{Heger2003}%
  \BibitemOpen
  \bibfield  {author} {\bibinfo {author} {\bibfnamefont {A.}~\bibnamefont
  {Heger}}, \bibinfo {author} {\bibfnamefont {C.~L.}\ \bibnamefont {Fryer}},
  \bibinfo {author} {\bibfnamefont {S.~E.}\ \bibnamefont {Woosley}}, \bibinfo
  {author} {\bibfnamefont {N.}~\bibnamefont {Langer}}, \ and\ \bibinfo {author}
  {\bibfnamefont {D.~H.}\ \bibnamefont {Hartmann}},\ }\href {\doibase
  10.1086/375341} {\bibfield  {journal} {\bibinfo  {journal} {Astrophys. J.}\
  }\textbf {\bibinfo {volume} {591}},\ \bibinfo {pages} {288} (\bibinfo {year}
  {2003})}\BibitemShut {NoStop}%
\bibitem [{\citenamefont {Dominik}\ \emph {et~al.}(2015)\citenamefont
  {Dominik}, \citenamefont {Berti}, \citenamefont {O'Shaughnessy},
  \citenamefont {Mandel}, \citenamefont {Belczynski}, \citenamefont {Fryer},
  \citenamefont {Holz}, \citenamefont {Bulik},\ and\ \citenamefont
  {Pannarale}}]{Dominik2015}%
  \BibitemOpen
  \bibfield  {author} {\bibinfo {author} {\bibfnamefont {M.}~\bibnamefont
  {Dominik}}, \bibinfo {author} {\bibfnamefont {E.}~\bibnamefont {Berti}},
  \bibinfo {author} {\bibfnamefont {R.}~\bibnamefont {O'Shaughnessy}}, \bibinfo
  {author} {\bibfnamefont {I.}~\bibnamefont {Mandel}}, \bibinfo {author}
  {\bibfnamefont {K.}~\bibnamefont {Belczynski}}, \bibinfo {author}
  {\bibfnamefont {C.}~\bibnamefont {Fryer}}, \bibinfo {author} {\bibfnamefont
  {D.~E.}\ \bibnamefont {Holz}}, \bibinfo {author} {\bibfnamefont
  {T.}~\bibnamefont {Bulik}}, \ and\ \bibinfo {author} {\bibfnamefont
  {F.}~\bibnamefont {Pannarale}},\ }\href {\doibase
  10.1088/0004-637X/806/2/263} {\bibfield  {journal} {\bibinfo  {journal}
  {Astrophys. J.}\ }\textbf {\bibinfo {volume} {806}},\ \bibinfo {pages} {263}
  (\bibinfo {year} {2015})}\BibitemShut {NoStop}%
\bibitem [{\citenamefont {Stevenson}\ \emph {et~al.}(2015)\citenamefont
  {Stevenson}, \citenamefont {Ohme},\ and\ \citenamefont
  {Fairhurst}}]{Stevenson2015}%
  \BibitemOpen
  \bibfield  {author} {\bibinfo {author} {\bibfnamefont {S.}~\bibnamefont
  {Stevenson}}, \bibinfo {author} {\bibfnamefont {F.}~\bibnamefont {Ohme}}, \
  and\ \bibinfo {author} {\bibfnamefont {S.}~\bibnamefont {Fairhurst}},\ }\href
  {\doibase 10.1088/0004-637X/810/1/58} {\bibfield  {journal} {\bibinfo
  {journal} {Astrophys. J.}\ }\textbf {\bibinfo {volume} {810}},\ \bibinfo
  {pages} {58} (\bibinfo {year} {2015})}\BibitemShut {NoStop}%
\bibitem [{\citenamefont {Belczynski}\ \emph
  {et~al.}(2016{\natexlab{b}})\citenamefont {Belczynski}, \citenamefont
  {Heger}, \citenamefont {Gladysz}, \citenamefont {Ruiter}, \citenamefont
  {Woosley}, \citenamefont {Wiktorowicz}, \citenamefont {Chen}, \citenamefont
  {Bulik}, \citenamefont {O'Shaughnesy}, \citenamefont {Holz}, \citenamefont
  {Fryer},\ and\ \citenamefont {Berti}}]{Belczynski2016a}%
  \BibitemOpen
  \bibfield  {author} {\bibinfo {author} {\bibfnamefont {K.}~\bibnamefont
  {Belczynski}}, \bibinfo {author} {\bibfnamefont {A.}~\bibnamefont {Heger}},
  \bibinfo {author} {\bibfnamefont {W.}~\bibnamefont {Gladysz}}, \bibinfo
  {author} {\bibfnamefont {A.~J.}\ \bibnamefont {Ruiter}}, \bibinfo {author}
  {\bibfnamefont {S.}~\bibnamefont {Woosley}}, \bibinfo {author} {\bibfnamefont
  {G.}~\bibnamefont {Wiktorowicz}}, \bibinfo {author} {\bibfnamefont {H.~Y.}\
  \bibnamefont {Chen}}, \bibinfo {author} {\bibfnamefont {T.}~\bibnamefont
  {Bulik}}, \bibinfo {author} {\bibfnamefont {R.}~\bibnamefont {O'Shaughnesy}},
  \bibinfo {author} {\bibfnamefont {D.~E.}\ \bibnamefont {Holz}}, \bibinfo
  {author} {\bibfnamefont {C.~L.}\ \bibnamefont {Fryer}}, \ and\ \bibinfo
  {author} {\bibfnamefont {E.}~\bibnamefont {Berti}},\ }\href {\doibase
  10.1051/0004-6361/201628980} {\bibfield  {journal} {\bibinfo  {journal}
  {Astron. Astrophys.}\ }\textbf {\bibinfo {volume} {594}},\ \bibinfo {pages}
  {1} (\bibinfo {year} {2016}{\natexlab{b}})}\BibitemShut {NoStop}%
\bibitem [{\citenamefont {Bartos}\ \emph {et~al.}(2017)\citenamefont {Bartos},
  \citenamefont {Haiman}, \citenamefont {Marka}, \citenamefont {Metzger},
  \citenamefont {Stone},\ and\ \citenamefont {Marka}}]{Bartos2017b}%
  \BibitemOpen
  \bibfield  {author} {\bibinfo {author} {\bibfnamefont {I.}~\bibnamefont
  {Bartos}}, \bibinfo {author} {\bibfnamefont {Z.}~\bibnamefont {Haiman}},
  \bibinfo {author} {\bibfnamefont {Z.}~\bibnamefont {Marka}}, \bibinfo
  {author} {\bibfnamefont {B.~D.}\ \bibnamefont {Metzger}}, \bibinfo {author}
  {\bibfnamefont {N.~C.}\ \bibnamefont {Stone}}, \ and\ \bibinfo {author}
  {\bibfnamefont {S.}~\bibnamefont {Marka}},\ }\href {\doibase
  10.1038/s41467-017-00851-7} {\bibfield  {journal} {\bibinfo  {journal} {Nat.
  Commun.}\ }\textbf {\bibinfo {volume} {8}},\ \bibinfo {pages} {831} (\bibinfo
  {year} {2017})}\BibitemShut {NoStop}%
\bibitem [{\citenamefont {Kovetz}\ \emph {et~al.}(2017)\citenamefont {Kovetz},
  \citenamefont {Cholis}, \citenamefont {Breysse},\ and\ \citenamefont
  {Kamionkowski}}]{Kovetz2017}%
  \BibitemOpen
  \bibfield  {author} {\bibinfo {author} {\bibfnamefont {E.~D.}\ \bibnamefont
  {Kovetz}}, \bibinfo {author} {\bibfnamefont {I.}~\bibnamefont {Cholis}},
  \bibinfo {author} {\bibfnamefont {P.~C.}\ \bibnamefont {Breysse}}, \ and\
  \bibinfo {author} {\bibfnamefont {M.}~\bibnamefont {Kamionkowski}},\ }\href
  {\doibase 10.1103/PhysRevD.95.103010} {\bibfield  {journal} {\bibinfo
  {journal} {Phys. Rev. D}\ }\textbf {\bibinfo {volume} {95}},\ \bibinfo
  {pages} {103010} (\bibinfo {year} {2017})}\BibitemShut {NoStop}%
\bibitem [{\citenamefont {Talbot}\ and\ \citenamefont
  {Thrane}(2017)}]{Talbot2017}%
  \BibitemOpen
  \bibfield  {author} {\bibinfo {author} {\bibfnamefont {C.}~\bibnamefont
  {Talbot}}\ and\ \bibinfo {author} {\bibfnamefont {E.}~\bibnamefont
  {Thrane}},\ }\href {\doibase 10.1103/PhysRevD.96.023012} {\bibfield
  {journal} {\bibinfo  {journal} {Phys. Rev. D}\ }\textbf {\bibinfo {volume}
  {96}},\ \bibinfo {pages} {023012} (\bibinfo {year} {2017})}\BibitemShut
  {NoStop}%
\bibitem [{\citenamefont {Wysocki}()}]{Wysocki2017}%
  \BibitemOpen
  \bibfield  {author} {\bibinfo {author} {\bibfnamefont {D.}~\bibnamefont
  {Wysocki}},\ }\href {http://arxiv.org/abs/1712.02643} {\ }\Eprint
  {http://arxiv.org/abs/1712.02643} {arXiv:1712.02643} \BibitemShut {NoStop}%
\bibitem [{\citenamefont {Miyamoto}\ \emph {et~al.}(2017)\citenamefont
  {Miyamoto}, \citenamefont {Kinugawa}, \citenamefont {Nakamura},\ and\
  \citenamefont {Kanda}}]{Miyamoto2017}%
  \BibitemOpen
  \bibfield  {author} {\bibinfo {author} {\bibfnamefont {A.}~\bibnamefont
  {Miyamoto}}, \bibinfo {author} {\bibfnamefont {T.}~\bibnamefont {Kinugawa}},
  \bibinfo {author} {\bibfnamefont {T.}~\bibnamefont {Nakamura}}, \ and\
  \bibinfo {author} {\bibfnamefont {N.}~\bibnamefont {Kanda}},\ }\href
  {\doibase 10.1103/PhysRevD.96.064025} {\bibfield  {journal} {\bibinfo
  {journal} {Phys. Rev. D}\ }\textbf {\bibinfo {volume} {96}},\ \bibinfo
  {pages} {064025} (\bibinfo {year} {2017})}\BibitemShut {NoStop}%
\bibitem [{\citenamefont {Mandel}\ \emph {et~al.}(2017)\citenamefont {Mandel},
  \citenamefont {Farr}, \citenamefont {Colonna}, \citenamefont {Stevenson},
  \citenamefont {Tiňo},\ and\ \citenamefont {Veitch}}]{Mandel2017}%
  \BibitemOpen
  \bibfield  {author} {\bibinfo {author} {\bibfnamefont {I.}~\bibnamefont
  {Mandel}}, \bibinfo {author} {\bibfnamefont {W.~M.}\ \bibnamefont {Farr}},
  \bibinfo {author} {\bibfnamefont {A.}~\bibnamefont {Colonna}}, \bibinfo
  {author} {\bibfnamefont {S.}~\bibnamefont {Stevenson}}, \bibinfo {author}
  {\bibfnamefont {P.}~\bibnamefont {Tiňo}}, \ and\ \bibinfo {author}
  {\bibfnamefont {J.}~\bibnamefont {Veitch}},\ }\href {\doibase
  10.1093/mnras/stw2883} {\bibfield  {journal} {\bibinfo  {journal} {Mon. Not.
  R. Astron. Soc.}\ }\textbf {\bibinfo {volume} {465}},\ \bibinfo {pages}
  {3254} (\bibinfo {year} {2017})}\BibitemShut {NoStop}%
\bibitem [{\citenamefont {Zevin}\ \emph {et~al.}(2017)\citenamefont {Zevin},
  \citenamefont {Pankow}, \citenamefont {Rodriguez}, \citenamefont {Sampson},
  \citenamefont {Chase}, \citenamefont {Kalogera},\ and\ \citenamefont
  {Rasio}}]{Zevin2017}%
  \BibitemOpen
  \bibfield  {author} {\bibinfo {author} {\bibfnamefont {M.}~\bibnamefont
  {Zevin}}, \bibinfo {author} {\bibfnamefont {C.}~\bibnamefont {Pankow}},
  \bibinfo {author} {\bibfnamefont {C.~L.}\ \bibnamefont {Rodriguez}}, \bibinfo
  {author} {\bibfnamefont {L.}~\bibnamefont {Sampson}}, \bibinfo {author}
  {\bibfnamefont {E.}~\bibnamefont {Chase}}, \bibinfo {author} {\bibfnamefont
  {V.}~\bibnamefont {Kalogera}}, \ and\ \bibinfo {author} {\bibfnamefont
  {F.~A.}\ \bibnamefont {Rasio}},\ }\href {\doibase 10.3847/1538-4357/aa8408}
  {\bibfield  {journal} {\bibinfo  {journal} {Astrophys. J.}\ }\textbf
  {\bibinfo {volume} {846}},\ \bibinfo {pages} {82} (\bibinfo {year}
  {2017})}\BibitemShut {NoStop}%
\bibitem [{\citenamefont {Stevenson}\ \emph
  {et~al.}(2017{\natexlab{a}})\citenamefont {Stevenson}, \citenamefont
  {Vigna-G{\'{o}}mez}, \citenamefont {Mandel}, \citenamefont {Barrett},
  \citenamefont {Neijssel}, \citenamefont {Perkins},\ and\ \citenamefont
  {de~Mink}}]{Stevenson2017}%
  \BibitemOpen
  \bibfield  {author} {\bibinfo {author} {\bibfnamefont {S.}~\bibnamefont
  {Stevenson}}, \bibinfo {author} {\bibfnamefont {A.}~\bibnamefont
  {Vigna-G{\'{o}}mez}}, \bibinfo {author} {\bibfnamefont {I.}~\bibnamefont
  {Mandel}}, \bibinfo {author} {\bibfnamefont {J.~W.}\ \bibnamefont {Barrett}},
  \bibinfo {author} {\bibfnamefont {C.~J.}\ \bibnamefont {Neijssel}}, \bibinfo
  {author} {\bibfnamefont {D.}~\bibnamefont {Perkins}}, \ and\ \bibinfo
  {author} {\bibfnamefont {S.~E.}\ \bibnamefont {de~Mink}},\ }\href {\doibase
  10.1038/ncomms14906} {\bibfield  {journal} {\bibinfo  {journal} {Nat.
  Commun.}\ }\textbf {\bibinfo {volume} {8}},\ \bibinfo {pages} {14906}
  (\bibinfo {year} {2017}{\natexlab{a}})}\BibitemShut {NoStop}%
\bibitem [{\citenamefont {Stevenson}\ \emph
  {et~al.}(2017{\natexlab{b}})\citenamefont {Stevenson}, \citenamefont
  {Berry},\ and\ \citenamefont {Mandel}}]{Stevenson2017a}%
  \BibitemOpen
  \bibfield  {author} {\bibinfo {author} {\bibfnamefont {S.}~\bibnamefont
  {Stevenson}}, \bibinfo {author} {\bibfnamefont {C.~P.~L.}\ \bibnamefont
  {Berry}}, \ and\ \bibinfo {author} {\bibfnamefont {I.}~\bibnamefont
  {Mandel}},\ }\href {\doibase 10.1093/mnras/stx1764} {\bibfield  {journal}
  {\bibinfo  {journal} {Mon. Not. R. Astron. Soc.}\ }\textbf {\bibinfo {volume}
  {2811}},\ \bibinfo {pages} {2801} (\bibinfo {year}
  {2017}{\natexlab{b}})}\BibitemShut {NoStop}%
\bibitem [{\citenamefont {{Barrett}}\ \emph {et~al.}(2018)\citenamefont
  {{Barrett}}, \citenamefont {{Gaebel}}, \citenamefont {{Neijssel}},
  \citenamefont {{Vigna-G{\'o}mez}}, \citenamefont {{Stevenson}}, \citenamefont
  {{Berry}}, \citenamefont {{Farr}},\ and\ \citenamefont
  {{Mandel}}}]{Barrett2018}%
  \BibitemOpen
  \bibfield  {author} {\bibinfo {author} {\bibfnamefont {J.~W.}\ \bibnamefont
  {{Barrett}}}, \bibinfo {author} {\bibfnamefont {S.~M.}\ \bibnamefont
  {{Gaebel}}}, \bibinfo {author} {\bibfnamefont {C.~J.}\ \bibnamefont
  {{Neijssel}}}, \bibinfo {author} {\bibfnamefont {A.}~\bibnamefont
  {{Vigna-G{\'o}mez}}}, \bibinfo {author} {\bibfnamefont {S.}~\bibnamefont
  {{Stevenson}}}, \bibinfo {author} {\bibfnamefont {C.~P.~L.}\ \bibnamefont
  {{Berry}}}, \bibinfo {author} {\bibfnamefont {W.~M.}\ \bibnamefont {{Farr}}},
  \ and\ \bibinfo {author} {\bibfnamefont {I.}~\bibnamefont {{Mandel}}},\
  }\href {\doibase 10.1093/mnras/sty908} {\bibfield  {journal} {\bibinfo
  {journal} {Mon. Not. R. Astron. Soc.}\ }\textbf {\bibinfo {volume} {477}},\
  \bibinfo {pages} {4685} (\bibinfo {year} {2018})}\BibitemShut {NoStop}%
\bibitem [{\citenamefont {Farr}\ \emph {et~al.}(2017)\citenamefont {Farr},
  \citenamefont {Stevenson}, \citenamefont {Miller}, \citenamefont {Mandel},
  \citenamefont {Farr},\ and\ \citenamefont {Vecchio}}]{Farr2017}%
  \BibitemOpen
  \bibfield  {author} {\bibinfo {author} {\bibfnamefont {W.~M.}\ \bibnamefont
  {Farr}}, \bibinfo {author} {\bibfnamefont {S.}~\bibnamefont {Stevenson}},
  \bibinfo {author} {\bibfnamefont {M.~C.}\ \bibnamefont {Miller}}, \bibinfo
  {author} {\bibfnamefont {I.}~\bibnamefont {Mandel}}, \bibinfo {author}
  {\bibfnamefont {B.}~\bibnamefont {Farr}}, \ and\ \bibinfo {author}
  {\bibfnamefont {A.}~\bibnamefont {Vecchio}},\ }\href {\doibase
  10.1038/nature23453} {\bibfield  {journal} {\bibinfo  {journal} {Nature
  (London)}\ }\textbf {\bibinfo {volume} {548}},\ \bibinfo {pages} {426}
  (\bibinfo {year} {2017})}\BibitemShut {NoStop}%
\bibitem [{\citenamefont {Fishbach}\ and\ \citenamefont
  {Holz}(2017)}]{Fishbach2017}%
  \BibitemOpen
  \bibfield  {author} {\bibinfo {author} {\bibfnamefont {M.}~\bibnamefont
  {Fishbach}}\ and\ \bibinfo {author} {\bibfnamefont {D.~E.}\ \bibnamefont
  {Holz}},\ }\href {\doibase 10.3847/2041-8213/aa9bf6} {\bibfield  {journal}
  {\bibinfo  {journal} {Astrophys. J. Lett.}\ }\textbf {\bibinfo {volume}
  {851}},\ \bibinfo {pages} {L25} (\bibinfo {year} {2017})}\BibitemShut
  {NoStop}%
\bibitem [{\citenamefont {O'Shaughnessy}\ \emph {et~al.}(2017)\citenamefont
  {O'Shaughnessy}, \citenamefont {Gerosa},\ and\ \citenamefont
  {Wysocki}}]{OShaughnessy2017}%
  \BibitemOpen
  \bibfield  {author} {\bibinfo {author} {\bibfnamefont {R.}~\bibnamefont
  {O'Shaughnessy}}, \bibinfo {author} {\bibfnamefont {D.}~\bibnamefont
  {Gerosa}}, \ and\ \bibinfo {author} {\bibfnamefont {D.}~\bibnamefont
  {Wysocki}},\ }\href {\doibase 10.1103/PhysRevLett.119.011101} {\bibfield
  {journal} {\bibinfo  {journal} {Phys. Rev. Lett.}\ }\textbf {\bibinfo
  {volume} {119}},\ \bibinfo {pages} {011101} (\bibinfo {year}
  {2017})}\BibitemShut {NoStop}%
\bibitem [{\citenamefont {Belczynski}\ \emph {et~al.}()\citenamefont
  {Belczynski}, \citenamefont {Klencki}, \citenamefont {Meynet}, \citenamefont
  {Fryer}, \citenamefont {Brown}, \citenamefont {Chruslinska}, \citenamefont
  {Gladysz}, \citenamefont {O'Shaughnessy}, \citenamefont {Bulik},
  \citenamefont {Berti}, \citenamefont {Holz}, \citenamefont {Gerosa},
  \citenamefont {Giersz}, \citenamefont {Ekstrom}, \citenamefont {Georgy},
  \citenamefont {Askar}, \citenamefont {Wysocki},\ and\ \citenamefont
  {Lasota}}]{Belczynski2017}%
  \BibitemOpen
  \bibfield  {author} {\bibinfo {author} {\bibfnamefont {K.}~\bibnamefont
  {Belczynski}}, \bibinfo {author} {\bibfnamefont {J.}~\bibnamefont {Klencki}},
  \bibinfo {author} {\bibfnamefont {G.}~\bibnamefont {Meynet}}, \bibinfo
  {author} {\bibfnamefont {C.~L.}\ \bibnamefont {Fryer}}, \bibinfo {author}
  {\bibfnamefont {D.~A.}\ \bibnamefont {Brown}}, \bibinfo {author}
  {\bibfnamefont {M.}~\bibnamefont {Chruslinska}}, \bibinfo {author}
  {\bibfnamefont {W.}~\bibnamefont {Gladysz}}, \bibinfo {author} {\bibfnamefont
  {R.}~\bibnamefont {O'Shaughnessy}}, \bibinfo {author} {\bibfnamefont
  {T.}~\bibnamefont {Bulik}}, \bibinfo {author} {\bibfnamefont
  {E.}~\bibnamefont {Berti}}, \bibinfo {author} {\bibfnamefont {D.~E.}\
  \bibnamefont {Holz}}, \bibinfo {author} {\bibfnamefont {D.}~\bibnamefont
  {Gerosa}}, \bibinfo {author} {\bibfnamefont {M.}~\bibnamefont {Giersz}},
  \bibinfo {author} {\bibfnamefont {S.}~\bibnamefont {Ekstrom}}, \bibinfo
  {author} {\bibfnamefont {C.}~\bibnamefont {Georgy}}, \bibinfo {author}
  {\bibfnamefont {A.}~\bibnamefont {Askar}}, \bibinfo {author} {\bibfnamefont
  {D.}~\bibnamefont {Wysocki}}, \ and\ \bibinfo {author} {\bibfnamefont
  {J.~P.}\ \bibnamefont {Lasota}},\ }\href {http://arxiv.org/abs/1706.07053} {\
  }\Eprint {http://arxiv.org/abs/1706.07053} {arXiv:1706.07053} \BibitemShut
  {NoStop}%
\bibitem [{\citenamefont {Wysocki}\ \emph {et~al.}(2018)\citenamefont
  {Wysocki}, \citenamefont {Gerosa}, \citenamefont {O'Shaughnessy},
  \citenamefont {Belczynski}, \citenamefont {Gladysz}, \citenamefont {Berti},
  \citenamefont {Kesden},\ and\ \citenamefont {Holz}}]{Wysocki2018}%
  \BibitemOpen
  \bibfield  {author} {\bibinfo {author} {\bibfnamefont {D.}~\bibnamefont
  {Wysocki}}, \bibinfo {author} {\bibfnamefont {D.}~\bibnamefont {Gerosa}},
  \bibinfo {author} {\bibfnamefont {R.}~\bibnamefont {O'Shaughnessy}}, \bibinfo
  {author} {\bibfnamefont {K.}~\bibnamefont {Belczynski}}, \bibinfo {author}
  {\bibfnamefont {W.}~\bibnamefont {Gladysz}}, \bibinfo {author} {\bibfnamefont
  {E.}~\bibnamefont {Berti}}, \bibinfo {author} {\bibfnamefont
  {M.}~\bibnamefont {Kesden}}, \ and\ \bibinfo {author} {\bibfnamefont {D.~E.}\
  \bibnamefont {Holz}},\ }\href {\doibase 10.1103/PhysRevD.97.043014}
  {\bibfield  {journal} {\bibinfo  {journal} {Phys. Rev. D}\ }\textbf {\bibinfo
  {volume} {97}},\ \bibinfo {pages} {043014} (\bibinfo {year}
  {2018})}\BibitemShut {NoStop}%
\bibitem [{\citenamefont {Talbot}\ and\ \citenamefont
  {Thrane}(2018)}]{Talbot2018}%
  \BibitemOpen
  \bibfield  {author} {\bibinfo {author} {\bibfnamefont {C.}~\bibnamefont
  {Talbot}}\ and\ \bibinfo {author} {\bibfnamefont {E.}~\bibnamefont
  {Thrane}},\ }\href {\doibase 10.3847/1538-4357/AAB34C} {\bibfield  {journal}
  {\bibinfo  {journal} {Astrophys. J.}\ }\textbf {\bibinfo {volume} {856}},\
  \bibinfo {pages} {173} (\bibinfo {year} {2018})}\BibitemShut {NoStop}%
\bibitem [{\citenamefont {Breivik}\ \emph {et~al.}(2016)\citenamefont
  {Breivik}, \citenamefont {Rodriguez}, \citenamefont {Larson}, \citenamefont
  {Kalogera},\ and\ \citenamefont {Rasio}}]{Breivik2016}%
  \BibitemOpen
  \bibfield  {author} {\bibinfo {author} {\bibfnamefont {K.}~\bibnamefont
  {Breivik}}, \bibinfo {author} {\bibfnamefont {C.~L.}\ \bibnamefont
  {Rodriguez}}, \bibinfo {author} {\bibfnamefont {S.~L.}\ \bibnamefont
  {Larson}}, \bibinfo {author} {\bibfnamefont {V.}~\bibnamefont {Kalogera}}, \
  and\ \bibinfo {author} {\bibfnamefont {F.~A.}\ \bibnamefont {Rasio}},\ }\href
  {\doibase 10.3847/2041-8205/830/1/L18} {\bibfield  {journal} {\bibinfo
  {journal} {Astrophys. J. Lett.}\ }\textbf {\bibinfo {volume} {830}},\
  \bibinfo {pages} {L18} (\bibinfo {year} {2016})}\BibitemShut {NoStop}%
\bibitem [{\citenamefont {Nishizawa}\ \emph {et~al.}(2016)\citenamefont
  {Nishizawa}, \citenamefont {Berti}, \citenamefont {Klein},\ and\
  \citenamefont {Sesana}}]{Nishizawa2016}%
  \BibitemOpen
  \bibfield  {author} {\bibinfo {author} {\bibfnamefont {A.}~\bibnamefont
  {Nishizawa}}, \bibinfo {author} {\bibfnamefont {E.}~\bibnamefont {Berti}},
  \bibinfo {author} {\bibfnamefont {A.}~\bibnamefont {Klein}}, \ and\ \bibinfo
  {author} {\bibfnamefont {A.}~\bibnamefont {Sesana}},\ }\href {\doibase
  10.1103/PhysRevD.94.064020} {\bibfield  {journal} {\bibinfo  {journal} {Phys.
  Rev. D}\ }\textbf {\bibinfo {volume} {94}},\ \bibinfo {pages} {064020}
  (\bibinfo {year} {2016})}\BibitemShut {NoStop}%
\bibitem [{\citenamefont {Sesana}(2016)}]{Sesana2016}%
  \BibitemOpen
  \bibfield  {author} {\bibinfo {author} {\bibfnamefont {A.}~\bibnamefont
  {Sesana}},\ }\href {\doibase 10.1103/PhysRevLett.116.231102} {\bibfield
  {journal} {\bibinfo  {journal} {Phys. Rev. Lett.}\ }\textbf {\bibinfo
  {volume} {116}},\ \bibinfo {pages} {231102} (\bibinfo {year}
  {2016})}\BibitemShut {NoStop}%
\bibitem [{\citenamefont {Mandel}\ and\ \citenamefont {{de
  Mink}}(2016)}]{Mandel2016}%
  \BibitemOpen
  \bibfield  {author} {\bibinfo {author} {\bibfnamefont {I.}~\bibnamefont
  {Mandel}}\ and\ \bibinfo {author} {\bibfnamefont {S.~E.}\ \bibnamefont {{de
  Mink}}},\ }\href {\doibase 10.1093/mnras/stw379} {\bibfield  {journal}
  {\bibinfo  {journal} {Mon. Not. R. Astron. Soc.}\ }\textbf {\bibinfo {volume}
  {458}},\ \bibinfo {pages} {2634} (\bibinfo {year} {2016})}\BibitemShut
  {NoStop}%
\bibitem [{\citenamefont {de~Mink}\ and\ \citenamefont
  {Mandel}(2016)}]{demink2016}%
  \BibitemOpen
  \bibfield  {author} {\bibinfo {author} {\bibfnamefont {S.~E.}\ \bibnamefont
  {de~Mink}}\ and\ \bibinfo {author} {\bibfnamefont {I.}~\bibnamefont
  {Mandel}},\ }\href {\doibase 10.1093/mnras/stw1219} {\bibfield  {journal}
  {\bibinfo  {journal} {Mon. Not. R. Astron. Soc.}\ }\textbf {\bibinfo {volume}
  {460}},\ \bibinfo {pages} {3545} (\bibinfo {year} {2016})}\BibitemShut
  {NoStop}%
\bibitem [{\citenamefont {{Tagawa}}\ \emph {et~al.}(2018)\citenamefont
  {{Tagawa}}, \citenamefont {{Saitoh}},\ and\ \citenamefont
  {{Kocsis}}}]{Tagawa2018}%
  \BibitemOpen
  \bibfield  {author} {\bibinfo {author} {\bibfnamefont {H.}~\bibnamefont
  {{Tagawa}}}, \bibinfo {author} {\bibfnamefont {T.~R.}\ \bibnamefont
  {{Saitoh}}}, \ and\ \bibinfo {author} {\bibfnamefont {B.}~\bibnamefont
  {{Kocsis}}},\ }\href {\doibase 10.1103/PhysRevLett.120.261101} {\bibfield
  {journal} {\bibinfo  {journal} {Phys. Rev. Lett.}\ }\textbf {\bibinfo
  {volume} {120}},\ \bibinfo {eid} {261101} (\bibinfo {year}
  {2018})}\BibitemShut {NoStop}%
\bibitem [{\citenamefont {Peters}(1964)}]{Peters1964}%
  \BibitemOpen
  \bibfield  {author} {\bibinfo {author} {\bibfnamefont {P.~C.}\ \bibnamefont
  {Peters}},\ }\href {\doibase 10.1103/PhysRev.136.B1224} {\bibfield  {journal}
  {\bibinfo  {journal} {Phys. Rev.}\ }\textbf {\bibinfo {volume} {136}},\
  \bibinfo {pages} {B1224} (\bibinfo {year} {1964})}\BibitemShut {NoStop}%
\bibitem [{\citenamefont {{Spitzer}}(1969)}]{Spitzer1969}%
  \BibitemOpen
  \bibfield  {author} {\bibinfo {author} {\bibfnamefont {L.}~\bibnamefont
  {{Spitzer}}, \bibfnamefont {Jr.}},\ }\href {\doibase 10.1086/180451}
  {\bibfield  {journal} {\bibinfo  {journal} {Astrophys. J. Lett}\ }\textbf
  {\bibinfo {volume} {158}},\ \bibinfo {pages} {L139} (\bibinfo {year}
  {1969})}\BibitemShut {NoStop}%
\bibitem [{\citenamefont {Freitag}\ \emph {et~al.}(2006)\citenamefont
  {Freitag}, \citenamefont {Amaro{-}Seoane},\ and\ \citenamefont
  {Kalogera}}]{Freitag2006}%
  \BibitemOpen
  \bibfield  {author} {\bibinfo {author} {\bibfnamefont {M.}~\bibnamefont
  {Freitag}}, \bibinfo {author} {\bibfnamefont {P.}~\bibnamefont
  {Amaro{-}Seoane}}, \ and\ \bibinfo {author} {\bibfnamefont {V.}~\bibnamefont
  {Kalogera}},\ }\href {\doibase 10.1086/506193} {\bibfield  {journal}
  {\bibinfo  {journal} {Astrophys. J.}\ }\textbf {\bibinfo {volume} {649}},\
  \bibinfo {pages} {91} (\bibinfo {year} {2006})}\BibitemShut {NoStop}%
\bibitem [{\citenamefont {Morscher}\ \emph {et~al.}(2013)\citenamefont
  {Morscher}, \citenamefont {Umbreit}, \citenamefont {Farr},\ and\
  \citenamefont {Rasio}}]{Morscher2013}%
  \BibitemOpen
  \bibfield  {author} {\bibinfo {author} {\bibfnamefont {M.}~\bibnamefont
  {Morscher}}, \bibinfo {author} {\bibfnamefont {S.}~\bibnamefont {Umbreit}},
  \bibinfo {author} {\bibfnamefont {W.~M.}\ \bibnamefont {Farr}}, \ and\
  \bibinfo {author} {\bibfnamefont {F.~A.}\ \bibnamefont {Rasio}},\ }\href
  {\doibase 10.1088/2041-8205/763/1/L15} {\bibfield  {journal} {\bibinfo
  {journal} {Astrophys. J. Lett.}\ }\textbf {\bibinfo {volume} {763}},\
  \bibinfo {pages} {2006} (\bibinfo {year} {2013})}\BibitemShut {NoStop}%
\bibitem [{\citenamefont {Samsing}\ \emph {et~al.}(2014)\citenamefont
  {Samsing}, \citenamefont {Macleod},\ and\ \citenamefont
  {Ramirez-Ruiz}}]{Samsing2014}%
  \BibitemOpen
  \bibfield  {author} {\bibinfo {author} {\bibfnamefont {J.}~\bibnamefont
  {Samsing}}, \bibinfo {author} {\bibfnamefont {M.}~\bibnamefont {Macleod}}, \
  and\ \bibinfo {author} {\bibfnamefont {E.}~\bibnamefont {Ramirez-Ruiz}},\
  }\href {\doibase 10.1088/0004-637X/784/1/71} {\bibfield  {journal} {\bibinfo
  {journal} {Astrophys. J.}\ }\textbf {\bibinfo {volume} {784}},\ \bibinfo
  {pages} {71} (\bibinfo {year} {2014})}\BibitemShut {NoStop}%
\bibitem [{\citenamefont {Rodriguez}\ \emph {et~al.}(2016)\citenamefont
  {Rodriguez}, \citenamefont {Chatterjee},\ and\ \citenamefont
  {Rasio}}]{Rodriguez2016ecc}%
  \BibitemOpen
  \bibfield  {author} {\bibinfo {author} {\bibfnamefont {C.~L.}\ \bibnamefont
  {Rodriguez}}, \bibinfo {author} {\bibfnamefont {S.}~\bibnamefont
  {Chatterjee}}, \ and\ \bibinfo {author} {\bibfnamefont {F.~A.}\ \bibnamefont
  {Rasio}},\ }\href {\doibase 10.1103/PhysRevD.93.084029} {\bibfield  {journal}
  {\bibinfo  {journal} {Phys. Rev. D}\ }\textbf {\bibinfo {volume} {93}},\
  \bibinfo {pages} {084029} (\bibinfo {year} {2016})}\BibitemShut {NoStop}%
\bibitem [{\citenamefont {Park}\ \emph {et~al.}(2017)\citenamefont {Park},
  \citenamefont {Kim}, \citenamefont {Lee}, \citenamefont {Bae},\ and\
  \citenamefont {Belczynski}}]{Park2017}%
  \BibitemOpen
  \bibfield  {author} {\bibinfo {author} {\bibfnamefont {D.}~\bibnamefont
  {Park}}, \bibinfo {author} {\bibfnamefont {C.}~\bibnamefont {Kim}}, \bibinfo
  {author} {\bibfnamefont {H.~M.}\ \bibnamefont {Lee}}, \bibinfo {author}
  {\bibfnamefont {Y.-B.}\ \bibnamefont {Bae}}, \ and\ \bibinfo {author}
  {\bibfnamefont {K.}~\bibnamefont {Belczynski}},\ }\href {\doibase
  10.1093/mnras/stx1015} {\bibfield  {journal} {\bibinfo  {journal} {Mon. Not.
  R. Astron. Soc.}\ }\textbf {\bibinfo {volume} {469}},\ \bibinfo {pages}
  {4665} (\bibinfo {year} {2017})}\BibitemShut {NoStop}%
\bibitem [{\citenamefont {Rodriguez}\ \emph {et~al.}(2018)\citenamefont
  {Rodriguez}, \citenamefont {Amaro-Seoane}, \citenamefont {Chatterjee},\ and\
  \citenamefont {Rasio}}]{Rodriguez2018a}%
  \BibitemOpen
  \bibfield  {author} {\bibinfo {author} {\bibfnamefont {C.~L.}\ \bibnamefont
  {Rodriguez}}, \bibinfo {author} {\bibfnamefont {P.}~\bibnamefont
  {Amaro-Seoane}}, \bibinfo {author} {\bibfnamefont {S.}~\bibnamefont
  {Chatterjee}}, \ and\ \bibinfo {author} {\bibfnamefont {F.~A.}\ \bibnamefont
  {Rasio}},\ }\href {\doibase 10.1103/PhysRevLett.120.151101} {\bibfield
  {journal} {\bibinfo  {journal} {Phys. Rev. Lett.}\ }\textbf {\bibinfo
  {volume} {120}},\ \bibinfo {pages} {151101} (\bibinfo {year}
  {2018})}\BibitemShut {NoStop}%
\bibitem [{\citenamefont {Samsing}(2018)}]{Samsing2018a}%
  \BibitemOpen
  \bibfield  {author} {\bibinfo {author} {\bibfnamefont {J.}~\bibnamefont
  {Samsing}},\ }\href {\doibase 10.1103/PhysRevD.97.103014} {\bibfield
  {journal} {\bibinfo  {journal} {Phys. Rev. D}\ }\textbf {\bibinfo {volume}
  {97}},\ \bibinfo {pages} {103014} (\bibinfo {year} {2018})}\BibitemShut
  {NoStop}%
\bibitem [{\citenamefont {{Samsing}}\ \emph {et~al.}(2018)\citenamefont
  {{Samsing}}, \citenamefont {{Askar}},\ and\ \citenamefont
  {{Giersz}}}]{Samsing2018b}%
  \BibitemOpen
  \bibfield  {author} {\bibinfo {author} {\bibfnamefont {J.}~\bibnamefont
  {{Samsing}}}, \bibinfo {author} {\bibfnamefont {A.}~\bibnamefont {{Askar}}},
  \ and\ \bibinfo {author} {\bibfnamefont {M.}~\bibnamefont {{Giersz}}},\
  }\href {\doibase 10.3847/1538-4357/aaab52} {\bibfield  {journal} {\bibinfo
  {journal} {{Astrophys. J.}}\ }\textbf {\bibinfo {volume} {855}},\ \bibinfo
  {eid} {124} (\bibinfo {year} {2018})}\BibitemShut {NoStop}%
\bibitem [{\citenamefont {{Samsing}}\ \emph {et~al.}()\citenamefont
  {{Samsing}}, \citenamefont {{D'Orazio}}, \citenamefont {{Askar}},\ and\
  \citenamefont {{Giersz}}}]{Samsing2018c}%
  \BibitemOpen
  \bibfield  {author} {\bibinfo {author} {\bibfnamefont {J.}~\bibnamefont
  {{Samsing}}}, \bibinfo {author} {\bibfnamefont {D.~J.}\ \bibnamefont
  {{D'Orazio}}}, \bibinfo {author} {\bibfnamefont {A.}~\bibnamefont {{Askar}}},
  \ and\ \bibinfo {author} {\bibfnamefont {M.}~\bibnamefont {{Giersz}}},\
  }\href@noop {} {\bibfield  {journal} {\bibinfo  {journal} {{Phys. Rev. D (to
  be published)}}\ }}\Eprint {http://arxiv.org/abs/1802.08654}
  {arXiv:1802.08654} \BibitemShut {NoStop}%
\bibitem [{\citenamefont {Abbott}\ \emph
  {et~al.}(2016{\natexlab{d}})\citenamefont {Abbott} \emph
  {et~al.}}]{150914prop}%
  \BibitemOpen
  \bibfield  {author} {\bibinfo {author} {\bibfnamefont {B.~P.}\ \bibnamefont
  {Abbott}} \emph {et~al.},\ }\href {\doibase 10.1103/PhysRevLett.116.241102}
  {\bibfield  {journal} {\bibinfo  {journal} {Phys. Rev. Lett.}\ }\textbf
  {\bibinfo {volume} {116}},\ \bibinfo {pages} {241102} (\bibinfo {year}
  {2016}{\natexlab{d}})}\BibitemShut {NoStop}%
\bibitem [{\citenamefont {{Abbott}}\ \emph {et~al.}(2017)\citenamefont
  {{Abbott}} \emph {et~al.}}]{150914modelsys}%
  \BibitemOpen
  \bibfield  {author} {\bibinfo {author} {\bibfnamefont {B.~P.}\ \bibnamefont
  {{Abbott}}} \emph {et~al.},\ }\href {\doibase 10.1088/1361-6382/aa6854}
  {\bibfield  {journal} {\bibinfo  {journal} {Classical Quantum Gravity}\
  }\textbf {\bibinfo {volume} {34}},\ \bibinfo {eid} {104002} (\bibinfo {year}
  {2017})}\BibitemShut {NoStop}%
\bibitem [{\citenamefont {Gond{\'a}n}\ \emph {et~al.}(2018)\citenamefont
  {Gond{\'a}n}, \citenamefont {Kocsis}, \citenamefont {Raffai},\ and\
  \citenamefont {Frei}}]{Gondan2018}%
  \BibitemOpen
  \bibfield  {author} {\bibinfo {author} {\bibfnamefont {L.}~\bibnamefont
  {Gond{\'a}n}}, \bibinfo {author} {\bibfnamefont {B.}~\bibnamefont {Kocsis}},
  \bibinfo {author} {\bibfnamefont {P.}~\bibnamefont {Raffai}}, \ and\ \bibinfo
  {author} {\bibfnamefont {Z.}~\bibnamefont {Frei}},\ }\href {\doibase
  https://doi.org/10.3847/1538-4357/aaad0e} {\bibfield  {journal} {\bibinfo
  {journal} {{Astrophys. J.}}\ }\textbf {\bibinfo {volume} {855}},\ \bibinfo
  {pages} {34} (\bibinfo {year} {2018})}\BibitemShut {NoStop}%
\bibitem [{\citenamefont {Vallisneri}(2008)}]{Vallisneri2008}%
  \BibitemOpen
  \bibfield  {author} {\bibinfo {author} {\bibfnamefont {M.}~\bibnamefont
  {Vallisneri}},\ }\href {\doibase 10.1103/PhysRevD.77.042001} {\bibfield
  {journal} {\bibinfo  {journal} {Phys. Rev. D}\ }\textbf {\bibinfo {volume}
  {77}},\ \bibinfo {pages} {042001} (\bibinfo {year} {2008})}\BibitemShut
  {NoStop}%
\bibitem [{\citenamefont {Ashton}\ \emph {et~al.}(shed)\citenamefont {Ashton},
  \citenamefont {H{\"u}bner}, \citenamefont {Lasky}, \citenamefont {Talbot},
  \citenamefont {Ackley}, \citenamefont {Biscoveanu}, \citenamefont {Chu},
  \citenamefont {Divarkala}, \citenamefont {Easter}, \citenamefont {Goncharov},
  \citenamefont {Hernandez}, \citenamefont {Harms}, \citenamefont {Lower},
  \citenamefont {Melchor}, \citenamefont {Payne}, \citenamefont {Pitkin},
  \citenamefont {Powell}, \citenamefont {Sarin}, \citenamefont {Smith},\ and\
  \citenamefont {Thrane}}]{BilbyInPrep}%
  \BibitemOpen
  \bibfield  {author} {\bibinfo {author} {\bibfnamefont {G.}~\bibnamefont
  {Ashton}}, \bibinfo {author} {\bibfnamefont {M.}~\bibnamefont {H{\"u}bner}},
  \bibinfo {author} {\bibfnamefont {P.~D.}\ \bibnamefont {Lasky}}, \bibinfo
  {author} {\bibfnamefont {C.}~\bibnamefont {Talbot}}, \bibinfo {author}
  {\bibfnamefont {K.}~\bibnamefont {Ackley}}, \bibinfo {author} {\bibfnamefont
  {S.~A.}\ \bibnamefont {Biscoveanu}}, \bibinfo {author} {\bibfnamefont
  {Q.}~\bibnamefont {Chu}}, \bibinfo {author} {\bibfnamefont {A.}~\bibnamefont
  {Divarkala}}, \bibinfo {author} {\bibfnamefont {P.}~\bibnamefont {Easter}},
  \bibinfo {author} {\bibfnamefont {B.}~\bibnamefont {Goncharov}}, \bibinfo
  {author} {\bibfnamefont {F.}~\bibnamefont {Hernandez}}, \bibinfo {author}
  {\bibfnamefont {J.}~\bibnamefont {Harms}}, \bibinfo {author} {\bibfnamefont
  {M.~E.}\ \bibnamefont {Lower}}, \bibinfo {author} {\bibfnamefont
  {D.}~\bibnamefont {Melchor}}, \bibinfo {author} {\bibfnamefont
  {E.}~\bibnamefont {Payne}}, \bibinfo {author} {\bibfnamefont {M.~D.}\
  \bibnamefont {Pitkin}}, \bibinfo {author} {\bibfnamefont {J.}~\bibnamefont
  {Powell}}, \bibinfo {author} {\bibfnamefont {N.}~\bibnamefont {Sarin}},
  \bibinfo {author} {\bibfnamefont {R.}~\bibnamefont {Smith}}, \ and\ \bibinfo
  {author} {\bibfnamefont {E.}~\bibnamefont {Thrane}},\ }\href@noop {} {\
  (\bibinfo {year} {to be published})}\BibitemShut {NoStop}%
\bibitem [{Note1()}]{Note1}%
  \BibitemOpen
  \bibinfo {note} {Examples of gravitational-wave injection and recovery with
  {\protect \sc Bilby} can be found here: \protect \url
  {https://git.ligo.org/lscsoft/bilby}}\BibitemShut {NoStop}%
\bibitem [{\citenamefont {{Buchner}}\ \emph {et~al.}(2014)\citenamefont
  {{Buchner}}, \citenamefont {{Georgakakis}}, \citenamefont {{Nandra}},
  \citenamefont {{Hsu}}, \citenamefont {{Rangel}}, \citenamefont {{Brightman}},
  \citenamefont {{Merloni}}, \citenamefont {{Salvato}}, \citenamefont
  {{Donley}},\ and\ \citenamefont {{Kocevski}}}]{PyMultiNest2014}%
  \BibitemOpen
  \bibfield  {author} {\bibinfo {author} {\bibfnamefont {J.}~\bibnamefont
  {{Buchner}}}, \bibinfo {author} {\bibfnamefont {A.}~\bibnamefont
  {{Georgakakis}}}, \bibinfo {author} {\bibfnamefont {K.}~\bibnamefont
  {{Nandra}}}, \bibinfo {author} {\bibfnamefont {L.}~\bibnamefont {{Hsu}}},
  \bibinfo {author} {\bibfnamefont {C.}~\bibnamefont {{Rangel}}}, \bibinfo
  {author} {\bibfnamefont {M.}~\bibnamefont {{Brightman}}}, \bibinfo {author}
  {\bibfnamefont {A.}~\bibnamefont {{Merloni}}}, \bibinfo {author}
  {\bibfnamefont {M.}~\bibnamefont {{Salvato}}}, \bibinfo {author}
  {\bibfnamefont {J.}~\bibnamefont {{Donley}}}, \ and\ \bibinfo {author}
  {\bibfnamefont {D.}~\bibnamefont {{Kocevski}}},\ }\href {\doibase
  10.1051/0004-6361/201322971} {\bibfield  {journal} {\bibinfo  {journal}
  {Astron. Astrophys.}\ }\textbf {\bibinfo {volume} {564}},\ \bibinfo {eid}
  {A125} (\bibinfo {year} {2014})}\BibitemShut {NoStop}%
\bibitem [{\citenamefont {{LIGO Scientific Collaboration}}()}]{LALSuite}%
  \BibitemOpen
  \bibfield  {author} {\bibinfo {author} {\bibnamefont {{LIGO Scientific
  Collaboration}}},\ }\href@noop {} {\enquote {\bibinfo {title} {{LIGO
  Scientific Collaboration Algorithm Library software packages: LALInference,
  LALSimulation, and LALInspiral}},}\ }\bibinfo {howpublished}
  {\url{https://wiki.ligo.org/Computing/DASWG/LALSuite}}\BibitemShut {NoStop}%
\bibitem [{\citenamefont {Hinder}\ \emph {et~al.}()\citenamefont {Hinder},
  \citenamefont {Kidder},\ and\ \citenamefont {Pfeiffer}}]{Hinder2017}%
  \BibitemOpen
  \bibfield  {author} {\bibinfo {author} {\bibfnamefont {I.}~\bibnamefont
  {Hinder}}, \bibinfo {author} {\bibfnamefont {L.~E.}\ \bibnamefont {Kidder}},
  \ and\ \bibinfo {author} {\bibfnamefont {H.~P.}\ \bibnamefont {Pfeiffer}},\
  }\href {\doibase 10.1103/PhysRevD.98.044015} {\bibfield  {journal} {\bibinfo
  {journal} {\prd}\ }\textbf {\bibinfo {volume} {98}},\ \bibinfo {pages}
  {044015}}\BibitemShut {NoStop}%
\bibitem [{\citenamefont {Huerta}\ \emph {et~al.}(2017)\citenamefont {Huerta},
  \citenamefont {Kumar}, \citenamefont {Agarwal}, \citenamefont {George},
  \citenamefont {Schive}, \citenamefont {Pfeiffer}, \citenamefont {Haas},
  \citenamefont {Ren}, \citenamefont {Chu}, \citenamefont {Boyle},
  \citenamefont {Hemberger}, \citenamefont {Kidder}, \citenamefont {Scheel},\
  and\ \citenamefont {Szilagyi}}]{Huerta2017a}%
  \BibitemOpen
  \bibfield  {author} {\bibinfo {author} {\bibfnamefont {E.~A.}\ \bibnamefont
  {Huerta}}, \bibinfo {author} {\bibfnamefont {P.}~\bibnamefont {Kumar}},
  \bibinfo {author} {\bibfnamefont {B.}~\bibnamefont {Agarwal}}, \bibinfo
  {author} {\bibfnamefont {D.}~\bibnamefont {George}}, \bibinfo {author}
  {\bibfnamefont {H.~Y.}\ \bibnamefont {Schive}}, \bibinfo {author}
  {\bibfnamefont {H.~P.}\ \bibnamefont {Pfeiffer}}, \bibinfo {author}
  {\bibfnamefont {R.}~\bibnamefont {Haas}}, \bibinfo {author} {\bibfnamefont
  {W.}~\bibnamefont {Ren}}, \bibinfo {author} {\bibfnamefont {T.}~\bibnamefont
  {Chu}}, \bibinfo {author} {\bibfnamefont {M.}~\bibnamefont {Boyle}}, \bibinfo
  {author} {\bibfnamefont {D.~A.}\ \bibnamefont {Hemberger}}, \bibinfo {author}
  {\bibfnamefont {L.~E.}\ \bibnamefont {Kidder}}, \bibinfo {author}
  {\bibfnamefont {M.~A.}\ \bibnamefont {Scheel}}, \ and\ \bibinfo {author}
  {\bibfnamefont {B.}~\bibnamefont {Szilagyi}},\ }\href {\doibase
  10.1103/PhysRevD.95.024038} {\bibfield  {journal} {\bibinfo  {journal} {Phys.
  Rev. D}\ }\textbf {\bibinfo {volume} {95}},\ \bibinfo {pages} {024038}
  (\bibinfo {year} {2017})}\BibitemShut {NoStop}%
\bibitem [{\citenamefont {Huerta}\ \emph {et~al.}(2014)\citenamefont {Huerta},
  \citenamefont {Kumar}, \citenamefont {McWilliams}, \citenamefont
  {O'Shaughnessy},\ and\ \citenamefont {Yunes}}]{Huerta2014}%
  \BibitemOpen
  \bibfield  {author} {\bibinfo {author} {\bibfnamefont {E.~A.}\ \bibnamefont
  {Huerta}}, \bibinfo {author} {\bibfnamefont {P.}~\bibnamefont {Kumar}},
  \bibinfo {author} {\bibfnamefont {S.~T.}\ \bibnamefont {McWilliams}},
  \bibinfo {author} {\bibfnamefont {R.}~\bibnamefont {O'Shaughnessy}}, \ and\
  \bibinfo {author} {\bibfnamefont {N.}~\bibnamefont {Yunes}},\ }\href
  {\doibase 10.1103/PhysRevD.90.084016} {\bibfield  {journal} {\bibinfo
  {journal} {Phys. Rev. D}\ }\textbf {\bibinfo {volume} {90}},\ \bibinfo
  {pages} {084016} (\bibinfo {year} {2014})}\BibitemShut {NoStop}%
\bibitem [{\citenamefont {Huerta}\ \emph {et~al.}(2018)\citenamefont {Huerta},
  \citenamefont {Moore}, \citenamefont {Kumar}, \citenamefont {George},
  \citenamefont {Chua}, \citenamefont {Haas}, \citenamefont {Wessel},
  \citenamefont {Johnson}, \citenamefont {Glennon}, \citenamefont {Rebei},
  \citenamefont {Holgado}, \citenamefont {Gair},\ and\ \citenamefont
  {Pfeiffer}}]{Huerta2018}%
  \BibitemOpen
  \bibfield  {author} {\bibinfo {author} {\bibfnamefont {E.~A.}\ \bibnamefont
  {Huerta}}, \bibinfo {author} {\bibfnamefont {C.~J.}\ \bibnamefont {Moore}},
  \bibinfo {author} {\bibfnamefont {P.}~\bibnamefont {Kumar}}, \bibinfo
  {author} {\bibfnamefont {D.}~\bibnamefont {George}}, \bibinfo {author}
  {\bibfnamefont {A.~J.~K.}\ \bibnamefont {Chua}}, \bibinfo {author}
  {\bibfnamefont {R.}~\bibnamefont {Haas}}, \bibinfo {author} {\bibfnamefont
  {E.}~\bibnamefont {Wessel}}, \bibinfo {author} {\bibfnamefont
  {D.}~\bibnamefont {Johnson}}, \bibinfo {author} {\bibfnamefont
  {D.}~\bibnamefont {Glennon}}, \bibinfo {author} {\bibfnamefont
  {A.}~\bibnamefont {Rebei}}, \bibinfo {author} {\bibfnamefont {A.~M.}\
  \bibnamefont {Holgado}}, \bibinfo {author} {\bibfnamefont {J.~R.}\
  \bibnamefont {Gair}}, \ and\ \bibinfo {author} {\bibfnamefont {H.~P.}\
  \bibnamefont {Pfeiffer}},\ }\href {\doibase
  http://dx.doi.org/10.1103/PhysRevD.97.024031} {\bibfield  {journal} {\bibinfo
   {journal} {Phys. Rev. D}\ }\textbf {\bibinfo {volume} {97}},\ \bibinfo
  {pages} {024031} (\bibinfo {year} {2018})}\BibitemShut {NoStop}%
\bibitem [{\citenamefont {Klein}\ \emph {et~al.}()\citenamefont {Klein},
  \citenamefont {Boetzel}, \citenamefont {Gopakumar}, \citenamefont {Jetzer},\
  and\ \citenamefont {de~Vittori}}]{Klein2018}%
  \BibitemOpen
  \bibfield  {author} {\bibinfo {author} {\bibfnamefont {A.}~\bibnamefont
  {Klein}}, \bibinfo {author} {\bibfnamefont {Y.}~\bibnamefont {Boetzel}},
  \bibinfo {author} {\bibfnamefont {A.}~\bibnamefont {Gopakumar}}, \bibinfo
  {author} {\bibfnamefont {P.}~\bibnamefont {Jetzer}}, \ and\ \bibinfo {author}
  {\bibfnamefont {L.}~\bibnamefont {de~Vittori}},\ }\href
  {http://arxiv.org/abs/1801.08542} {\bibfield  {journal} {\bibinfo  {journal}
  {Phys. Rev. D (to be published)}\ }}\Eprint {http://arxiv.org/abs/1801.08542}
  {arXiv:1801.08542} \BibitemShut {NoStop}%
\bibitem [{\citenamefont {Rebei}\ \emph {et~al.}()\citenamefont {Rebei},
  \citenamefont {Huerta}, \citenamefont {Wang}, \citenamefont {Habib},
  \citenamefont {Haas}, \citenamefont {Johnson},\ and\ \citenamefont
  {George}}]{Rebei18}%
  \BibitemOpen
  \bibfield  {author} {\bibinfo {author} {\bibfnamefont {A.}~\bibnamefont
  {Rebei}}, \bibinfo {author} {\bibfnamefont {E.~A.}\ \bibnamefont {Huerta}},
  \bibinfo {author} {\bibfnamefont {S.}~\bibnamefont {Wang}}, \bibinfo {author}
  {\bibfnamefont {S.}~\bibnamefont {Habib}}, \bibinfo {author} {\bibfnamefont
  {R.}~\bibnamefont {Haas}}, \bibinfo {author} {\bibfnamefont {D.}~\bibnamefont
  {Johnson}}, \ and\ \bibinfo {author} {\bibfnamefont {D.}~\bibnamefont
  {George}},\ }\href {http://arxiv.org/abs/1807.09787} {\bibfield  {journal}
  {\bibinfo  {journal} {Phys. Rev. Lett. (to be published)}\ }}\Eprint
  {http://arxiv.org/abs/1807.09787} {arXiv:1807.09787} \BibitemShut {NoStop}%
\bibitem [{\citenamefont {Mandel}\ \emph {et~al.}(2014)\citenamefont {Mandel},
  \citenamefont {Berry}, \citenamefont {Ohme}, \citenamefont {Fairhurst},\ and\
  \citenamefont {Farr}}]{Mandel2014}%
  \BibitemOpen
  \bibfield  {author} {\bibinfo {author} {\bibfnamefont {I.}~\bibnamefont
  {Mandel}}, \bibinfo {author} {\bibfnamefont {C.~P.~L.}\ \bibnamefont
  {Berry}}, \bibinfo {author} {\bibfnamefont {F.}~\bibnamefont {Ohme}},
  \bibinfo {author} {\bibfnamefont {S.}~\bibnamefont {Fairhurst}}, \ and\
  \bibinfo {author} {\bibfnamefont {W.~M.}\ \bibnamefont {Farr}},\ }\href
  {\doibase 10.1088/0264-9381/31/15/155005} {\bibfield  {journal} {\bibinfo
  {journal} {Classical Quantum Gravity}\ }\textbf {\bibinfo {volume} {31}}
  (\bibinfo {year} {2014}),\ 10.1088/0264-9381/31/15/155005}\BibitemShut
  {NoStop}%
\bibitem [{\citenamefont {{Messick}}\ \emph {et~al.}(2017)\citenamefont
  {{Messick}} \emph {et~al.}}]{gstlal}%
  \BibitemOpen
  \bibfield  {author} {\bibinfo {author} {\bibfnamefont {C.}~\bibnamefont
  {{Messick}}} \emph {et~al.},\ }\href {\doibase 10.1103/PhysRevD.95.042001}
  {\bibfield  {journal} {\bibinfo  {journal} {\prd}\ }\textbf {\bibinfo
  {volume} {95}},\ \bibinfo {eid} {042001} (\bibinfo {year}
  {2017})}\BibitemShut {NoStop}%
\bibitem [{\citenamefont {{Usman}}\ \emph {et~al.}(2016)\citenamefont {{Usman}}
  \emph {et~al.}}]{pycbc}%
  \BibitemOpen
  \bibfield  {author} {\bibinfo {author} {\bibfnamefont {S.~A.}\ \bibnamefont
  {{Usman}}} \emph {et~al.},\ }\href {\doibase 10.1088/0264-9381/33/21/215004}
  {\bibfield  {journal} {\bibinfo  {journal} {Classical Quantum Gravity}\
  }\textbf {\bibinfo {volume} {33}},\ \bibinfo {eid} {215004} (\bibinfo {year}
  {2016})}\BibitemShut {NoStop}%
\bibitem [{\citenamefont {{Tai}}\ \emph {et~al.}(2014)\citenamefont {{Tai}},
  \citenamefont {{McWilliams}},\ and\ \citenamefont {{Pretorius}}}]{Tai2014}%
  \BibitemOpen
  \bibfield  {author} {\bibinfo {author} {\bibfnamefont {K.~S.}\ \bibnamefont
  {{Tai}}}, \bibinfo {author} {\bibfnamefont {S.~T.}\ \bibnamefont
  {{McWilliams}}}, \ and\ \bibinfo {author} {\bibfnamefont {F.}~\bibnamefont
  {{Pretorius}}},\ }\href {\doibase 10.1103/PhysRevD.90.103001} {\bibfield
  {journal} {\bibinfo  {journal} {\prd}\ }\textbf {\bibinfo {volume} {90}},\
  \bibinfo {eid} {103001} (\bibinfo {year} {2014})}\BibitemShut {NoStop}%
\bibitem [{\citenamefont {{Coughlin}}\ \emph {et~al.}(2015)\citenamefont
  {{Coughlin}}, \citenamefont {{Meyers}}, \citenamefont {{Thrane}},
  \citenamefont {{Luo}},\ and\ \citenamefont {{Christensen}}}]{Coughlin2015}%
  \BibitemOpen
  \bibfield  {author} {\bibinfo {author} {\bibfnamefont {M.}~\bibnamefont
  {{Coughlin}}}, \bibinfo {author} {\bibfnamefont {P.}~\bibnamefont
  {{Meyers}}}, \bibinfo {author} {\bibfnamefont {E.}~\bibnamefont {{Thrane}}},
  \bibinfo {author} {\bibfnamefont {J.}~\bibnamefont {{Luo}}}, \ and\ \bibinfo
  {author} {\bibfnamefont {N.}~\bibnamefont {{Christensen}}},\ }\href {\doibase
  10.1103/PhysRevD.91.063004} {\bibfield  {journal} {\bibinfo  {journal}
  {\prd}\ }\textbf {\bibinfo {volume} {91}},\ \bibinfo {eid} {063004} (\bibinfo
  {year} {2015})}\BibitemShut {NoStop}%
\bibitem [{\citenamefont {{Tiwari}}\ \emph {et~al.}(2016)\citenamefont
  {{Tiwari}}, \citenamefont {{Klimenko}}, \citenamefont {{Christensen}},
  \citenamefont {{Huerta}}, \citenamefont {{Mohapatra}}, \citenamefont
  {{Gopakumar}}, \citenamefont {{Haney}}, \citenamefont {{Ajith}},
  \citenamefont {{McWilliams}}, \citenamefont {{Vedovato}}, \citenamefont
  {{Drago}}, \citenamefont {{Salemi}}, \citenamefont {{Prodi}}, \citenamefont
  {{Lazzaro}}, \citenamefont {{Tiwari}}, \citenamefont {{Mitselmakher}},\ and\
  \citenamefont {{Da Silva}}}]{Tiwari2016}%
  \BibitemOpen
  \bibfield  {author} {\bibinfo {author} {\bibfnamefont {V.}~\bibnamefont
  {{Tiwari}}}, \bibinfo {author} {\bibfnamefont {S.}~\bibnamefont
  {{Klimenko}}}, \bibinfo {author} {\bibfnamefont {N.}~\bibnamefont
  {{Christensen}}}, \bibinfo {author} {\bibfnamefont {E.~A.}\ \bibnamefont
  {{Huerta}}}, \bibinfo {author} {\bibfnamefont {S.~R.~P.}\ \bibnamefont
  {{Mohapatra}}}, \bibinfo {author} {\bibfnamefont {A.}~\bibnamefont
  {{Gopakumar}}}, \bibinfo {author} {\bibfnamefont {M.}~\bibnamefont
  {{Haney}}}, \bibinfo {author} {\bibfnamefont {P.}~\bibnamefont {{Ajith}}},
  \bibinfo {author} {\bibfnamefont {S.~T.}\ \bibnamefont {{McWilliams}}},
  \bibinfo {author} {\bibfnamefont {G.}~\bibnamefont {{Vedovato}}}, \bibinfo
  {author} {\bibfnamefont {M.}~\bibnamefont {{Drago}}}, \bibinfo {author}
  {\bibfnamefont {F.}~\bibnamefont {{Salemi}}}, \bibinfo {author}
  {\bibfnamefont {G.~A.}\ \bibnamefont {{Prodi}}}, \bibinfo {author}
  {\bibfnamefont {C.}~\bibnamefont {{Lazzaro}}}, \bibinfo {author}
  {\bibfnamefont {S.}~\bibnamefont {{Tiwari}}}, \bibinfo {author}
  {\bibfnamefont {G.}~\bibnamefont {{Mitselmakher}}}, \ and\ \bibinfo {author}
  {\bibfnamefont {F.}~\bibnamefont {{Da Silva}}},\ }\href {\doibase
  10.1103/PhysRevD.93.043007} {\bibfield  {journal} {\bibinfo  {journal}
  {\prd}\ }\textbf {\bibinfo {volume} {93}},\ \bibinfo {eid} {043007} (\bibinfo
  {year} {2016})}\BibitemShut {NoStop}%
\bibitem [{\citenamefont {{Brown}}\ and\ \citenamefont
  {{Zimmerman}}(2010)}]{Brown2010}%
  \BibitemOpen
  \bibfield  {author} {\bibinfo {author} {\bibfnamefont {D.~A.}\ \bibnamefont
  {{Brown}}}\ and\ \bibinfo {author} {\bibfnamefont {P.~J.}\ \bibnamefont
  {{Zimmerman}}},\ }\href {\doibase 10.1103/PhysRevD.81.024007} {\bibfield
  {journal} {\bibinfo  {journal} {\prd}\ }\textbf {\bibinfo {volume} {81}},\
  \bibinfo {eid} {024007} (\bibinfo {year} {2010})}\BibitemShut {NoStop}%
\bibitem [{\citenamefont {{Huerta}}\ and\ \citenamefont
  {{Brown}}(2013)}]{Huerta2013}%
  \BibitemOpen
  \bibfield  {author} {\bibinfo {author} {\bibfnamefont {E.~A.}\ \bibnamefont
  {{Huerta}}}\ and\ \bibinfo {author} {\bibfnamefont {D.~A.}\ \bibnamefont
  {{Brown}}},\ }\href {\doibase 10.1103/PhysRevD.87.127501} {\bibfield
  {journal} {\bibinfo  {journal} {\prd}\ }\textbf {\bibinfo {volume} {87}},\
  \bibinfo {eid} {127501} (\bibinfo {year} {2013})}\BibitemShut {NoStop}%
\bibitem [{\citenamefont {Abbott}\ \emph
  {et~al.}(2016{\natexlab{e}})\citenamefont {Abbott} \emph
  {et~al.}}]{150914improved}%
  \BibitemOpen
  \bibfield  {author} {\bibinfo {author} {\bibfnamefont {B.~P.}\ \bibnamefont
  {Abbott}} \emph {et~al.},\ }\href {\doibase 10.1103/PhysRevX.6.041014}
  {\bibfield  {journal} {\bibinfo  {journal} {Phys. Rev. X}\ }\textbf {\bibinfo
  {volume} {6}},\ \bibinfo {pages} {041014} (\bibinfo {year}
  {2016}{\natexlab{e}})}\BibitemShut {NoStop}%
\bibitem [{\citenamefont {{Advanced LIGO anticipated sensitivity
  curves}}()}]{LIGOpsd}%
  \BibitemOpen
  \bibfield  {author} {\bibinfo {author} {\bibnamefont {{Advanced LIGO
  anticipated sensitivity curves}}},\ }\href@noop {} {}\bibinfo {howpublished}
  {\url{https://dcc.ligo.org/LIGO-T0900288/public}}\BibitemShut {NoStop}%
\bibitem [{\citenamefont {{Prospects for Observing and Localizing
  Gravitational-Wave Transients with Advanced LIGO, Advanced Virgo and
  KAGRA}}()}]{Virgopsd}%
  \BibitemOpen
  \bibfield  {author} {\bibinfo {author} {\bibnamefont {{Prospects for
  Observing and Localizing Gravitational-Wave Transients with Advanced LIGO,
  Advanced Virgo and KAGRA}}},\ }\href@noop {} {}\bibinfo {howpublished}
  {\url{https://dcc.ligo.org/LIGO-P1200087-v42/public}}\BibitemShut {NoStop}%
\bibitem [{\citenamefont {Flanagan}\ and\ \citenamefont
  {Hughes}(1998)}]{Flanagan1998}%
  \BibitemOpen
  \bibfield  {author} {\bibinfo {author} {\bibfnamefont {{\'E}.~{\'E}.}\
  \bibnamefont {Flanagan}}\ and\ \bibinfo {author} {\bibfnamefont {S.~A.}\
  \bibnamefont {Hughes}},\ }\href {\doibase 10.1103/PhysRevD.57.4566}
  {\bibfield  {journal} {\bibinfo  {journal} {Phys. Rev. D}\ }\textbf {\bibinfo
  {volume} {57}},\ \bibinfo {pages} {4566} (\bibinfo {year}
  {1998})}\BibitemShut {NoStop}%
\bibitem [{\citenamefont {Lindblom}\ \emph {et~al.}(2008)\citenamefont
  {Lindblom}, \citenamefont {Owen},\ and\ \citenamefont
  {Brown}}]{Lindblom2008}%
  \BibitemOpen
  \bibfield  {author} {\bibinfo {author} {\bibfnamefont {L.}~\bibnamefont
  {Lindblom}}, \bibinfo {author} {\bibfnamefont {B.~J.}\ \bibnamefont {Owen}},
  \ and\ \bibinfo {author} {\bibfnamefont {D.~A.}\ \bibnamefont {Brown}},\
  }\href {\doibase 10.1103/PhysRevD.78.124020} {\bibfield  {journal} {\bibinfo
  {journal} {Phys. Rev. D}\ }\textbf {\bibinfo {volume} {78}},\ \bibinfo
  {pages} {124020} (\bibinfo {year} {2008})}\BibitemShut {NoStop}%
\bibitem [{\citenamefont {Baird}\ \emph {et~al.}(2013)\citenamefont {Baird},
  \citenamefont {Fairhurst}, \citenamefont {Hannam},\ and\ \citenamefont
  {Murphy}}]{Baird}%
  \BibitemOpen
  \bibfield  {author} {\bibinfo {author} {\bibfnamefont {E.}~\bibnamefont
  {Baird}}, \bibinfo {author} {\bibfnamefont {S.}~\bibnamefont {Fairhurst}},
  \bibinfo {author} {\bibfnamefont {M.}~\bibnamefont {Hannam}}, \ and\ \bibinfo
  {author} {\bibfnamefont {P.}~\bibnamefont {Murphy}},\ }\href@noop {}
  {\bibfield  {journal} {\bibinfo  {journal} {Phys. Rev. D}\ }\textbf {\bibinfo
  {volume} {87}},\ \bibinfo {pages} {024035} (\bibinfo {year}
  {2013})}\BibitemShut {NoStop}%
\bibitem [{\citenamefont {Abbott}\ \emph {et~al.}(2017)\citenamefont {Abbott}
  \emph {et~al.}}]{CosmicExplorer}%
  \BibitemOpen
  \bibfield  {author} {\bibinfo {author} {\bibfnamefont {B.~P.}\ \bibnamefont
  {Abbott}} \emph {et~al.},\ }\href {\doibase 10.1088/1361-6382/aa51f4}
  {\bibfield  {journal} {\bibinfo  {journal} {Classical Quantum Gravity}\
  }\textbf {\bibinfo {volume} {34}},\ \bibinfo {pages} {44001} (\bibinfo {year}
  {2017})}\BibitemShut {NoStop}%
\bibitem [{\citenamefont {Punturo}\ \emph {et~al.}(2010)\citenamefont {Punturo}
  \emph {et~al.}}]{EinsteinTelescope}%
  \BibitemOpen
  \bibfield  {author} {\bibinfo {author} {\bibfnamefont {M.}~\bibnamefont
  {Punturo}} \emph {et~al.},\ }\href
  {http://stacks.iop.org/0264-9381/27/i=19/a=194002} {\bibfield  {journal}
  {\bibinfo  {journal} {Classical Quantum Gravity}\ }\textbf {\bibinfo {volume}
  {27}},\ \bibinfo {pages} {194002} (\bibinfo {year} {2010})}\BibitemShut
  {NoStop}%
\bibitem [{\citenamefont {{Exploring the Sensitivity of Next Generation
  Gravitational Wave Detectors}}()}]{CEpsd}%
  \BibitemOpen
  \bibfield  {author} {\bibinfo {author} {\bibnamefont {{Exploring the
  Sensitivity of Next Generation Gravitational Wave Detectors}}},\ }\href@noop
  {} {}\bibinfo {howpublished}
  {\url{https://dcc.ligo.org/LIGO-P1600143/public}}\BibitemShut {NoStop}%
\bibitem [{\citenamefont {{ET Sensitivities Page}}()}]{ETpsd}%
  \BibitemOpen
  \bibfield  {author} {\bibinfo {author} {\bibnamefont {{ET Sensitivities
  Page}}},\ }\href@noop {} {}\bibinfo {howpublished}
  {\url{http://www.et-gw.eu/index.php/etsensitivities}}\BibitemShut {NoStop}%
\bibitem [{\citenamefont {Hild}\ \emph {et~al.}(2011)\citenamefont {Hild} \emph
  {et~al.}}]{Hild2011}%
  \BibitemOpen
  \bibfield  {author} {\bibinfo {author} {\bibfnamefont {S.}~\bibnamefont
  {Hild}} \emph {et~al.},\ }\href {\doibase 10.1088/0264-9381/28/9/094013}
  {\bibfield  {journal} {\bibinfo  {journal} {Classical Quantum Gravity}\
  }\textbf {\bibinfo {volume} {28}},\ \bibinfo {pages} {094013} (\bibinfo
  {year} {2011})}\BibitemShut {NoStop}%
\bibitem [{\citenamefont {O'leary}\ \emph {et~al.}(2009)\citenamefont
  {O'leary}, \citenamefont {Kocsis},\ and\ \citenamefont {Loeb}}]{Oleary2009}%
  \BibitemOpen
  \bibfield  {author} {\bibinfo {author} {\bibfnamefont {R.~M.}\ \bibnamefont
  {O'leary}}, \bibinfo {author} {\bibfnamefont {B.}~\bibnamefont {Kocsis}}, \
  and\ \bibinfo {author} {\bibfnamefont {A.}~\bibnamefont {Loeb}},\ }\href
  {\doibase 10.1111/j.1365-2966.2009.14653.x} {\bibfield  {journal} {\bibinfo
  {journal} {Mon. Not. R. Astron. Soc.}\ }\textbf {\bibinfo {volume} {395}},\
  \bibinfo {pages} {2127} (\bibinfo {year} {2009})}\BibitemShut {NoStop}%
\bibitem [{\citenamefont {Tsang}(2013)}]{Tsang2013}%
  \BibitemOpen
  \bibfield  {author} {\bibinfo {author} {\bibfnamefont {D.}~\bibnamefont
  {Tsang}},\ }\href {\doibase 10.1088/0004-637X/777/2/103} {\bibfield
  {journal} {\bibinfo  {journal} {Astrophys. J.}\ }\textbf {\bibinfo {volume}
  {777}},\ \bibinfo {pages} {103} (\bibinfo {year} {2013})}\BibitemShut
  {NoStop}%
\bibitem [{\citenamefont {{Antonini}}\ and\ \citenamefont
  {{Perets}}(2012)}]{Antonini2012}%
  \BibitemOpen
  \bibfield  {author} {\bibinfo {author} {\bibfnamefont {F.}~\bibnamefont
  {{Antonini}}}\ and\ \bibinfo {author} {\bibfnamefont {H.~B.}\ \bibnamefont
  {{Perets}}},\ }\href {\doibase 10.1088/0004-637X/757/1/27} {\bibfield
  {journal} {\bibinfo  {journal} {{Astrophys. J.}}\ }\textbf {\bibinfo {volume}
  {757}},\ \bibinfo {eid} {27} (\bibinfo {year} {2012})}\BibitemShut {NoStop}%
\bibitem [{\citenamefont {{Hoang}}\ \emph {et~al.}(2018)\citenamefont
  {{Hoang}}, \citenamefont {{Naoz}}, \citenamefont {{Kocsis}}, \citenamefont
  {{Rasio}},\ and\ \citenamefont {{Dosopoulou}}}]{Hoang2018}%
  \BibitemOpen
  \bibfield  {author} {\bibinfo {author} {\bibfnamefont {B.-M.}\ \bibnamefont
  {{Hoang}}}, \bibinfo {author} {\bibfnamefont {S.}~\bibnamefont {{Naoz}}},
  \bibinfo {author} {\bibfnamefont {B.}~\bibnamefont {{Kocsis}}}, \bibinfo
  {author} {\bibfnamefont {F.~A.}\ \bibnamefont {{Rasio}}}, \ and\ \bibinfo
  {author} {\bibfnamefont {F.}~\bibnamefont {{Dosopoulou}}},\ }\href {\doibase
  10.3847/1538-4357/aaafce} {\bibfield  {journal} {\bibinfo  {journal}
  {{Astrophys. J.}}\ }\textbf {\bibinfo {volume} {856}},\ \bibinfo {eid} {140}
  (\bibinfo {year} {2018})}\BibitemShut {NoStop}%
\bibitem [{\citenamefont {{Hamers}}\ \emph {et~al.}(2018)\citenamefont
  {{Hamers}}, \citenamefont {{Bar-Or}}, \citenamefont {{Petrovich}},\ and\
  \citenamefont {{Antonini}}}]{Hamers2018}%
  \BibitemOpen
  \bibfield  {author} {\bibinfo {author} {\bibfnamefont {A.~S.}\ \bibnamefont
  {{Hamers}}}, \bibinfo {author} {\bibfnamefont {B.}~\bibnamefont {{Bar-Or}}},
  \bibinfo {author} {\bibfnamefont {C.}~\bibnamefont {{Petrovich}}}, \ and\
  \bibinfo {author} {\bibfnamefont {F.}~\bibnamefont {{Antonini}}},\ }\href
  {\doibase 10.3847/1538-4357/aadae2} {\bibfield  {journal} {\bibinfo
  {journal} {Astrophys. J.}\ }\textbf {\bibinfo {volume} {863}},\ \bibinfo
  {pages} {7} (\bibinfo {year} {2018})}\BibitemShut {NoStop}%
\bibitem [{\citenamefont {Antonini}\ \emph {et~al.}(2017)\citenamefont
  {Antonini}, \citenamefont {Toonen},\ and\ \citenamefont
  {Hamers}}]{Antonini2017}%
  \BibitemOpen
  \bibfield  {author} {\bibinfo {author} {\bibfnamefont {F.}~\bibnamefont
  {Antonini}}, \bibinfo {author} {\bibfnamefont {S.}~\bibnamefont {Toonen}}, \
  and\ \bibinfo {author} {\bibfnamefont {A.~S.}\ \bibnamefont {Hamers}},\
  }\href {\doibase 10.3847/1538-4357/aa6f5e} {\bibfield  {journal} {\bibinfo
  {journal} {Astrophys. J.}\ }\textbf {\bibinfo {volume} {841}},\ \bibinfo
  {pages} {77} (\bibinfo {year} {2017})}\BibitemShut {NoStop}%
\bibitem [{\citenamefont {Silsbee}\ and\ \citenamefont
  {Tremaine}(2017)}]{Silsbee2017}%
  \BibitemOpen
  \bibfield  {author} {\bibinfo {author} {\bibfnamefont {K.}~\bibnamefont
  {Silsbee}}\ and\ \bibinfo {author} {\bibfnamefont {S.}~\bibnamefont
  {Tremaine}},\ }\href {\doibase 10.3847/1538-4357/aa5729} {\bibfield
  {journal} {\bibinfo  {journal} {Astrophys. J.}\ }\textbf {\bibinfo {volume}
  {836}},\ \bibinfo {pages} {39} (\bibinfo {year} {2017})}\BibitemShut
  {NoStop}%
\bibitem [{\citenamefont {Rodriguez}\ and\ \citenamefont
  {Antonini}(2018)}]{Rodriguez2018b}%
  \BibitemOpen
  \bibfield  {author} {\bibinfo {author} {\bibfnamefont {C.~L.}\ \bibnamefont
  {Rodriguez}}\ and\ \bibinfo {author} {\bibfnamefont {F.}~\bibnamefont
  {Antonini}},\ }\href {\doibase 10.3847/1538-4357/aacea4} {\bibfield
  {journal} {\bibinfo  {journal} {Astrophys. J.}\ }\textbf {\bibinfo {volume}
  {863}},\ \bibinfo {pages} {7} (\bibinfo {year} {2018})}\BibitemShut {NoStop}%
\bibitem [{\citenamefont {Kocsis}\ and\ \citenamefont
  {Levin}(2012)}]{Kocsis2012}%
  \BibitemOpen
  \bibfield  {author} {\bibinfo {author} {\bibfnamefont {B.}~\bibnamefont
  {Kocsis}}\ and\ \bibinfo {author} {\bibfnamefont {J.}~\bibnamefont {Levin}},\
  }\href {\doibase 10.1103/PhysRevD.85.123005} {\bibfield  {journal} {\bibinfo
  {journal} {Phys. Rev. D}\ }\textbf {\bibinfo {volume} {85}},\ \bibinfo
  {pages} {123005} (\bibinfo {year} {2012})}\BibitemShut {NoStop}%
\bibitem [{\citenamefont {Wen}(2003)}]{Wen2003}%
  \BibitemOpen
  \bibfield  {author} {\bibinfo {author} {\bibfnamefont {L.}~\bibnamefont
  {Wen}},\ }\href {\doibase 10.1086/378794} {\bibfield  {journal} {\bibinfo
  {journal} {Astrophys. J.}\ }\textbf {\bibinfo {volume} {598}},\ \bibinfo
  {pages} {419} (\bibinfo {year} {2003})}\BibitemShut {NoStop}%
\bibitem [{\citenamefont {Hinder}\ \emph {et~al.}(2008)\citenamefont {Hinder},
  \citenamefont {Vaishnav}, \citenamefont {Herrmann}, \citenamefont
  {Shoemaker},\ and\ \citenamefont {Laguna}}]{Hinder2008}%
  \BibitemOpen
  \bibfield  {author} {\bibinfo {author} {\bibfnamefont {I.}~\bibnamefont
  {Hinder}}, \bibinfo {author} {\bibfnamefont {B.}~\bibnamefont {Vaishnav}},
  \bibinfo {author} {\bibfnamefont {F.}~\bibnamefont {Herrmann}}, \bibinfo
  {author} {\bibfnamefont {D.~M.}\ \bibnamefont {Shoemaker}}, \ and\ \bibinfo
  {author} {\bibfnamefont {P.}~\bibnamefont {Laguna}},\ }\href {\doibase
  10.1103/PhysRevD.77.081502} {\bibfield  {journal} {\bibinfo  {journal} {Phys.
  Rev. D}\ }\textbf {\bibinfo {volume} {77}},\ \bibinfo {pages} {081502}
  (\bibinfo {year} {2008})}\BibitemShut {NoStop}%
\end{thebibliography}%

\end{document}